\documentclass{aa}
\usepackage{pdflscape}
\usepackage{graphicx}
\usepackage{txfonts}
\usepackage{textcomp, gensymb}
\usepackage{lscape}
\usepackage[dvipsnames]{xcolor}
\usepackage{placeins}
\usepackage[version=4]{mhchem} 
\usepackage[colorlinks = true,
            linkcolor = blue,
            urlcolor  = blue,
            citecolor = blue,
            anchorcolor = blue, unicode]{hyperref}

\begin{document}

   \title{\ce{H2O} emission as tracer of pebble drift: insights from coupling transport and thermochemical models}
   \titlerunning{\ce{H2O} emission as tracer of pebble drift}

   \author{Marissa Vlasblom\inst{1}
          \and
          Andrew D. Sellek\inst{1}
          \and
          Ewine F. van Dishoeck\inst{1,2}
          }

   \institute{Leiden Observatory, Leiden University, 2300 RA Leiden, Netherlands\\ 
              \email{vlasblom@strw.leidenuniv.nl}
    \and
    Max-Planck Institut f\"{u}r Extraterrestrische Physik (MPE), Giessenbachstr. 1, 85748, Garching, Germany 
   }
   \date{Received xxx; accepted yyy}
 
  \abstract 
   {The composition of the inner regions of protoplanetary disks is known to change with time due to the delivery of icy grains. Cold \ce{H2O} emission is often hypothesized to be a tracer of this pebble drift, as sublimating \ce{H2O} ices may be mixed vertically upwards into the surface layers of the disk, where this cold gas may be observed with JWST. However, it is unclear to what extent processes such as photodissociation or the co-delivery of dust, which would obscure the delivered \ce{H2O}, may impede such detections.}
   {The transport of gas and dust through a disk is often considered with 1D models, whereas more complex 2D models that include a full chemical network are often static in nature. We aim to combine the two modeling approaches to obtain an improved, 2D view of transport in disks, to better understand how these processes can be traced by, for example, \ce{H2O} emission as seen with JWST-MIRI. }
   {We combine the 1D transport code DiscEvolution with the 2D thermochemical code DALI to create several grids of models in which the gas-phase abundances, dust properties, or both are varied according to the transport model. Several snapshots from a DiscEvolution are taken and post-processed in DALI, considering times from $10^4$ to $4\times10^6$ yr. Delivered abundances are mixed up vertically instantaneously, after which the chemistry is run for $10^3$ yr. We consider one scenario in which a traffic jam is present inside the \ce{H2O} snowline, allowing dust to pile up there, and one scenario in which this traffic jam is not present.}
   {The transport of both gas and dust leads to significant temperature changes within the disk, which strongly influence line fluxes and ratios. When a traffic jam is present, the delivery of \ce{H2O} can indeed proceed unnoticed due to the co-delivery of dust, as the retrieved \ce{H2O} mass from LTE slab fits does not change. In addition, the relative strength of cold \ce{H2O} lines is not found to be sensitive to this delivery of gas-phase \ce{H2O}: it would need to rapidly diffuse outwards and upwards in the disk after sublimation to be observed as such with JWST-MIRI. Instead, the cold \ce{H2O} lines can be greatly enhanced by the delivery of only dust to the inner disk when a traffic jam is present, due to the preferential obscuration of the warm and hot \ce{H2O} reservoirs by this dust pile-up. Therefore, one can create a spectrum with strong cold \ce{H2O} emission (a signature hypothesized to trace the delivery of \ce{H2O} ices across the snowline) solely through the delivery of dust rather than \ce{H2O}. The case of \ce{CO2}/\ce{H2O} as a tracer is briefly discussed as well, but without any clear trend found. }
   {Recent work has used the 1500/6000 K \ce{H2O} line ratio as a proxy to determine the pebble mass flux crossing the \ce{H2O} snowline, but we find that the use of this line ratio for this purpose warrants some caution. The line ratio is rather sensitive to the temperature and dust distribution of the disk, and is often influenced not only by the cold \ce{H2O} mass, but also by the hot \ce{H2O} mass. Both of these can change considerably during the evolution of the disk, which can introduce complexities and trends as a function of time that do not match the true evolution of the pebble flux with time. }


   \maketitle
%
\nolinenumbers
\section{Introduction} \label{sec:intro}

Young, Earth-like planets are thought to form primarily in the inner regions ($<$10 au) of protoplanetary disks, and therefore the chemical composition of their atmospheres and that of the disk should be closely linked \citep{dawson2018, oberg2021}. However, the composition of the inner disk is expected not to be static with time. Dust grains experience inward radial drift \citep{weidenschilling1977}, and they carry ices that desorb into the gas-phase at their respective ice lines. Therefore, the chemical make-up of the gas in the inner regions of the disk will change with time. In particular, this has been shown to affect the carbon-to-oxygen (C/O) ratio of the gas, as well as the O/H and C/H ratios \citep[e.g.][]{booth2017, mah2023, sellek2025_carbon, molyarova2026}. 

The enhancement of the inner disk gas by sublimating ices occurs approximately in a sequence, set by the respective binding energies of the ices. \ce{H2O} ice reaches the inner disk first, as it has a relatively high binding energy and therefore a snowline located close to the star \citep[e.g.][]{minissale2022}. When this ice desorbs in the inner disk, it significantly increases the O/H abundance and thereby lowers the C/O ratio. This is followed by the arrival of \ce{CO2} ices, which have a slightly lower binding energy and therefore travel slightly farther from their ice line to reach the inner disk. The \ce{CO2} ices cause the C/O ratio to start to rise slightly. Over time, the oxygen-rich gas will drain from the inner disk onto the central star, after which it will gradually be replenished instead with advected carbon-rich gas that contains mainly CO and \ce{CH4}, two major ice components with a low binding energy. Hence, at late times, the C/O ratio will continue to rise further, potentially even reaching values above unity \citep{mah2023, sellek2025_carbon}. 

The inner regions of the protoplanetary disk are best studied in the infrared (IR), as their warm temperatures allow for the excitation of molecular ro-vibrational lines in the mid-IR from species such as \ce{H2O}, as well as \ce{CO2}, CO, \ce{C2H2} and HCN \citep[e.g.][]{pontoppidan2014}. The Mid-InfraRed Instrument on board the \textit{James Webb} Space Telescope \citep[JWST-MIRI;][]{rieke2015, wright2015, wright2023} has allowed for the chemical composition of these regions to be characterized in further detail than before \citep[see, e.g.][]{henning2024, vandishoeck2023, arulanantham2025}, with higher sensitivity and spectral resolution than previous space-based instruments. 
The increased spectral resolution has allowed for the \ce{H2O} emission in disks to be characterized in greater detail than before, where lines of different upper level energies $E_{\rm up}$ can now be used to identify distinct temperature reservoirs \citep[see, e.g.][]{gasman2023b, temmink2024b, temmink2025, romero-mirza2024_sample, banzatti2023, banzatti2025}. Additionally, the increased sensitivity has motivated the hunt for less-abundant species and rare isotopologues, including a tentative detection of \ce{H2^18O} \citep{salyk2026}.

One major question that recent works have been trying to answer is whether the aforementioned transport of icy pebbles into the inner disk leaves traces in the IR emission coming from this region. In particular, the \ce{H2O} emission has been studied extensively, with recent works proposing that cold ($E_{\rm up} \lesssim 2000$ K) \ce{H2O} lines may hold some information about the mass flux of icy pebbles crossing the \ce{H2O} snowline \citep{banzatti2023, banzatti2025, romero-mirza2024_sample, krijt2025}. This is motivated by an empirical correlation between the cold \ce{H2O} strength and the dust disk size as measured by ALMA, where smaller, more compact disks (which may be caused by efficient radial drift) show stronger cold \ce{H2O}. LTE slab modeling suggests that these lines trace a cold reservoir with temperatures of $\sim$200 K, close to the sublimation temperature of \ce{H2O}. When a large flux of icy pebbles crosses the snowline, the resulting sublimation has been hypothesized to enhance this cold reservoir, more than the warmer \ce{H2O} reservoirs that trace temperatures $>$400 K. This is then thought to potentially translate into an enhancement in flux of cold, $E_{\rm up} \sim 1500$ K \ce{H2O} lines with respect to warmer, $E_{\rm up} \gtrsim 3000$ K lines \citep[e.g.,][]{banzatti2023, temmink2024b, temmink2025, vlasblom2025_H2O}. 

However, the feasibility of observing the delivery of \ce{H2O} into the inner disk by pebble drift has also been called into question by recent modeling work. \citet{sellek2025_CO2} and \citet{houge2025_H2O} point out that the delivery of gas-phase \ce{H2O} comes paired with the delivery of dust grains, which increase the dust opacity in the IR emitting layers of the disk. Therefore, the \ce{H2O} gas could essentially be ``smuggled unnoticed'' into the inner disk, as the increase in column density by sublimating ices and the increase in dust opacity by dust grains may very well (partially) cancel one another out. 

Additionally, \citet{houge2025_H2O} and \citet{vlasblom2025_H2O} demonstrate the importance of vertical mixing and photodissociation in the IR-emitting layers of the disk. \citet{vlasblom2025_H2O} demonstrate using 2D thermochemical models how the cold \ce{H2O} lines thought to trace pebble drift are emitted from a layer above the snow surface at radii beyond 1 au (i.e., larger radii than the 0.5 au midplane snowline). Thus, these lines do not trace the midplane of the disk, the region where pebble drift truly occurs. However, \citet{vlasblom2025_H2O} demonstrate that disk chemistry alone is incapable of reproducing typical observed line ratios; the \ce{H2O} in this layer is photodissociated rapidly, yet its abundance needs to be enhanced by several orders of magnitude. Potentially, this could be achieved through the sublimation of \ce{H2O} gas from icy pebbles followed by quickly mixing upwards vertically (and also radially outwards), thus establishing a correlation of the line strength with the pebble flux.
However, photodissociation timescales in this surface layer are fast ($\lesssim$1 yr; e.g. \citealt{xu2019, kanwar2024_model}), and therefore whether this enhancement by sublimating ices can last long enough to be observed could be debated. However, if a continuous supply is present and a sufficient column has built up, \ce{H2O} UV-shielding may help \citep{bethell2009, bosman2022a}. 

Therefore, the transport of gas and icy dust grains in protoplanetary disks should be considered in more than just the inward radial direction, as processes such as photodissociation, UV shielding, and chemical reactions may be important, especially in the surface layers which are observed with JWST. However, current work typically considers the transport of gas and dust in disk models only in 1D \citep[e.g.][]{sellek2025_CO2, mah2023, houge2025_H2O}. On the other hand, more complex models that self-consistently calculate the 2D chemical abundances in the disk using a full chemical network are typically static in nature and do not consider mixing or changing abundances with time \citep[e.g.][]{vlasblom2025_H2O, arabhavi2026}. Some efforts have been made to combine the two. Most notably, \citet{greenwood2019_dustevol} combined the thermochemical code {\tt ProDiMo} \citep{woitke2009, woitke2024} with the dust evolution model {\tt two-pop-py} \citep{birnstiel_2017_twopoppy} to demonstrate how IR fluxes of commonly detected species with \textit{Spitzer} changed with time as the dust in the disk evolved. However, they considered only the changes this would bring in the dust composition of the disk, not the elemental composition. Turbulent mixing processes have been implemented into chemical models before \citep{semenov2006_turbulence, semenov2011}, and more recently, \citet{woitke2022} implemented vertical mixing and diffusion processes into {\tt ProDiMo}, but did not link this to any physical model of dust and gas transport.  

In this work, we present a study that combines the 1D transport code DiscEvolution with the 2D thermochemical code DALI such that the effects of both dust and gas transport can be studied with further complexity. In Sect. \ref{sec:methods}, we present both codes and our approach to coupling them. In Sect. \ref{sec:results}, the effects of gas and dust transport on the abundance and emission of \ce{H2O} is presented. In Sect. \ref{sec:discussion}, we discuss the implications for the use of cold \ce{H2O} lines as a tracer of pebble drift. We also discuss whether the emission from \ce{CO2}, another prominent ice species, may give further insights. Sect. \ref{sec:conclusions} summarizes our main conclusions. 


\section{Methods}\label{sec:methods}

This work combines the 1D transport code DiscEvolution \citep{booth2017, sellek2025_CO2, booth2026_DiscEvolution} and the 2D thermochemical code DALI \citep{bruderer2012, bruderer2013} to obtain a 2D view of the effects of gas and dust transport to the inner disk. In this section, we introduce both codes and describe how they are linked. 

\subsection{1D transport code DiscEvolution}\label{subsec:methods_1d}

The advection of gas and transport of (icy) dust is modeled using DiscEvolution. The code is set up as described in \citet{sellek2025_CO2}, so we refer the reader to that work for the full description and only summarize the most important details here. 


DiscEvolution adopts the two-population dust model from \citet{birnstiel2012}. The dust consists of a population of small grains, which is assumed to have a fixed size, and follows the motion of the gas. The population of large grains, on the other hand, is allowed to grow in size up to either a fragmentation barrier, when grains collide at high-enough velocities such that they can no longer grow larger efficiently, or up to a drift barrier, when grains are removed from the disk by radial drift faster than they can grow. This latter population can thus also dynamically decouple from the gas once they grow large enough. The fraction of dust mass in each population is given by $f_{\rm small}$ and $f_{\rm large}$, which are bench-marked against fragmentation-coagulation simulations \citep{birnstiel2012}. These suggest that $f_{\rm large} = 0.75$ when growth is fragmentation-limited, and $f_{\rm large} = 0.97$ when growth is drift-limited. 

For this work, we run two models with DiscEvolution. The first model is equal to Scenario 1 presented in \citet{sellek2025_CO2}. In this model, both the dust and gas evolve according to a viscosity of $\alpha=10^{-3}$, and a `traffic jam' is present: icy grains are assumed to have a higher fragmentation velocity than bare grains ($u_{\rm f, dust}=1$ m s$^{-1}$ and $u_{\rm f, ice}=10$ m s$^{-1}$). This leads to a pile-up of dust inside the \ce{H2O} snowline \citep[see also][]{pinilla2016}, as grains in this region will have a lower Stokes number and a lower velocity, thus becoming more coupled to the gas. We also run a model where this traffic jam effect is not present. In this case, both the gas and dust still evolve according to a viscosity of $\alpha=10^{-3}$, but the fragmentation velocity is now set to $u_{\rm f, ice}=10$ m s$^{-1}$ throughout the entire disk. This value is higher than recent laboratory experiments suggest \citep[e.g.][]{gundlach2018, musiolik2019}, however this allows us to retain the same degree of coupling between the gas and dust in the outer disk for the same evolutionary timescales of the gas. This ensures that radial drift, and the subsequent sublimation of ices at the snowlines, still occurs at the same rate and is unaffected by this change. 

In Appendix \ref{app:figures}, Fig. \ref{fig:1Dcode_abu} presents the combined gas- and ice-phase abundances of the four major species we evolve in this work: \ce{H2O, CO2, CH4}, and CO, as a function of disk radius at several time steps. This demonstrates how the (gas-phase) abundance of each species first builds up locally just within its snowline due to ice sublimation, after which this gas is viscously advected inwards leading to a uniform enhancement within its snowline. Fig. \ref{fig:1Dcode_abu} also presents the dust surface density at several time steps for the two scenarios (with and without a traffic jam) considered in this work.

\subsection{2D Thermochemical code DALI}\label{subsec:methods_DALI}

The thermochemical code DALI consists of three main steps, first calculating the local UV and IR radiation field and dust temperature at all locations in the disk given an input gas and dust density structure. Then, the code self-consistently calculates the gas temperature, chemical abundances of all species in the chemical network, and non-LTE excitation for several specified atoms and molecules (which is then used in the heating and cooling balance). Finally, a raytracing tool can be used to obtain line fluxes, spectra and spectral image cubes. 

All DALI models have a general setup similar to those presented in \citet{vlasblom2024, vlasblom2025_H2O}, as they are based on the model for the disk around K5 star AS 209 \citep[see][]{zhang2021, bosman2022a}. Hence, our models use the {observed} stellar spectrum from those works {as input}. {This spectrum} has an effective temperature of 4300 K and a bolometric luminosity of 1.4 $L_\odot$. {It also contains an FUV component due to its accretion, with an FUV luminosity of $8\times10^{-3}\; L_\odot$, which corresponds to a UV-excess produced by an accretion rate between $10^{-8}-10^{-9}$ M$_\odot$ yr$^{-1}$}. 

The output from our DiscEvolution simulations is then transferred to DALI. This is done by taking the DiscEvolution output at several time steps and transforming each snapshot from the code into a single DALI model. We consider 6 logarithmically spaced time steps between $10^4$ and $4\times10^6$ years (see colored lines in Fig. \ref{fig:1Dcode_abu}), as well as the initial $t=0$ condition (black line in Fig. \ref{fig:1Dcode_abu}). {The DiscEvolution models start with an initial disk mass of 0.028 M$_\odot$, which has decreased to 0.021 M$_\odot$ due to viscous evolution by the latest timestep we consider. However, since we aim to isolate the effects of changing chemical abundances and dust properties, and this mass decrease is relatively small, we fix the total disk mass in all DALI models to 0.021 M$_\odot$.}

\begin{table*}[]
    \centering
    \caption{Summary of all grids presented in this work.}
    \begin{tabular}{l l l}
        \hline
        \hline
         Grid name & Abundances & Dust properties \\
         \hline
         Abu. only & DiscEvolution & Fiducial - $\Delta_{\rm gas/dust} = 100$, $f_{\rm large} = 0.9$, and $\chi = 0.2$ \\
         Dust only (Traffic jam) & Fiducial atomic abundances & DiscEvolution - $u_{\rm f, dust}=1$ m s$^{-1}$, $u_{\rm f, ice}=10$ m s$^{-1}$ \\
         Dust only (No traffic jam) & Fiducial atomic abundances & DiscEvolution - $u_{\rm f, dust}=u_{\rm f, ice}=10$ m s$^{-1}$ \\
         Abu.+dust (Traffic jam) & DiscEvolution & DiscEvolution - $u_{\rm f, dust}=1$ m s$^{-1}$, $u_{\rm f, ice}=10$ m s$^{-1}$ \\
         Abu.+dust (No traffic jam) & DiscEvolution & DiscEvolution - $u_{\rm f, dust}=u_{\rm f, ice}=10$ m s$^{-1}$ \\
         \hline
         
    \end{tabular}
    \label{tab:grids}
\end{table*}

We transfer two kinds of output from DiscEvolution to DALI: the chemical abundances and the dust properties. To test the effects of each separately, we create five grids of models (each grid consisting of 7 models corresponding to our 7 time steps). We create one grid using a fiducial DALI setup for the dust properties \citep[see, e.g.][]{vlasblom2025_H2O}, where the chemical abundances are varied according to the DiscEvolution output. We also create two grids where the chemical abundances are kept constant at a fiducial value, and the dust properties are varied according to our two DiscEvolution simulations (with and without a traffic jam). Finally, we create two grids in which both quantities are varied simultaneously. A summary of all grids presented in this work and their properties is shown in Table \ref{tab:grids}.

\begin{figure*}
    \centering
    \includegraphics[width=\linewidth]{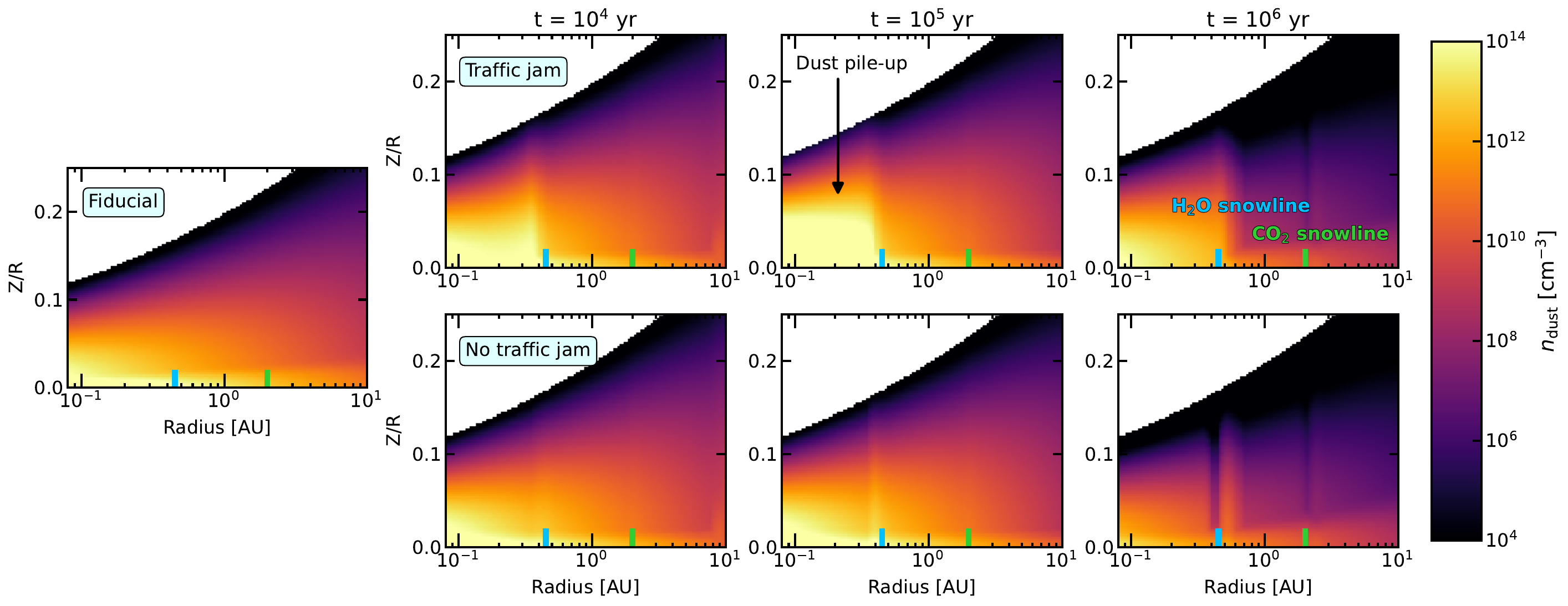}
    \caption{Dust density distribution (small + large grains) given by DALI. The leftmost panel shows the fiducial distribution used by the grid with varied abundances only (see Table \ref{tab:grids}). The six remaining panels show the dust density created from the output of DiscEvolution at three different time steps, in both the scenarios with (top row) and without a traffic jam (bottom row).}
    \label{fig:ndust}
\end{figure*}

In DALI, the dust distribution is modeled using two populations, similar to DiscEvolution. The small dust population follows the same vertical distribution as the gas \citep{miotello2016}, and has an MRN size distribution \citep{mathis1977} with grain sizes between 5 nm and 1 $\mu$m. The large dust population has an MRN size distribution with grain sizes between 5 nm and 1 mm, and has a reduced scale height compared to the small dust population. {Though slight differences may exist, we assume that the two dust populations from the two codes directly map to one another. Each dust population has its own fixed dust opacity curve for a standard ISM dust composition \citep[consistent with][]{andrews2011, bruderer2012}. These are kept the same for all models presented in this work, and thus do not account for changes in dust properties with time.} {We assume} a gas surface density profile for a viscously evolving disk \citep{lyndenbell1974, hartmann1998}
\begin{align}
    \Sigma_{\rm gas} = \Sigma_{\rm c} \left( \frac{R}{R_{\rm c}} \right)^{-\gamma} \exp \left[ - \left( \frac{R}{R_{\rm c}}\right)^{2-\gamma} \right], \label{eq:sigma_gas}
\end{align}
{and this profile is kept fixed for all models, since it is not expected to change significantly over the timescale considered by this work.} The dust densities {are changed for each model, and } the two distributions are thus parameterized as follows:
\begin{align}
    \rho_{\rm dust,small} &= \frac{(1-f_{\rm large}) \Sigma_{\rm gas} \Delta_{\rm gas/dust}}{\sqrt{2\pi}Rh} \exp\left[ -\frac{1}{2}\left( \frac{\pi/2-\theta}{h} \right)^2 \right], \label{eq:small_dust}\\
    \rho_{\rm dust,large} &= \frac{f_{\rm large} \Sigma_{\rm gas} \Delta_{\rm gas/dust}}{\sqrt{2\pi}R\chi h} \exp\left[ -\frac{1}{2}\left( \frac{\pi/2-\theta}{\chi h} \right)^2 \right].\label{eq:large_dust}
\end{align}

Here, $R = \sqrt{r^2+z^2}$, $\theta=\arctan(r/z)$, and $h = h_{\rm c}(R/R_{\rm c})^\psi$. We assume $R_{\rm c} = 46$ au, {$\Sigma_{\rm c} = 21.32$ g cm$^{-2}$}, $h_{\rm c} = 0.08$, $\gamma = 1$, and $\psi = 0.25$. At each radius $r$ in the disk, the dust density as a function of height is therefore determined according to the vertically integrated gas/dust ratio ($\Delta_{\rm gas/dust}$) and the midplane values of the mass fraction in large grains ($f_{\rm large}$) and settling factor ($\chi$). All three of {these} are given by DiscEvolution {and thus change both with time and location in the disk in response to the grain size changes predicted by the two-population model of \citet{birnstiel2012}}. The resulting 2D dust distribution is presented in Fig. \ref{fig:ndust} for several time steps, both for the scenario with and without a traffic jam. As a comparison, a fiducial dust distribution is also given (which is used in the grid where only the abundances are varied). In this model, we assume $\Delta_{\rm gas/dust} = 100$, $f_{\rm large} = 0.9$, and $\chi = 0.2$ throughout the entire disk \citep[which are the same as in][]{vlasblom2025_H2O}. In Appendix \ref{app:figures}, 
{Fig. \ref{fig:1Dcode_dustprops} presents the dust properties as a function of radius as modeled in DiscEvolution ($\Delta_{\rm gas/dust}$, $f_{\rm large}$, and $\chi$) for the scenarios with and without a traffic jam. }
Fig. \ref{fig:gasdust} presents the {resulting 2D map of the} gas-to-dust ratio throughout the disk.

Finally, the abundances from \ce{H2O, CO2, CH4} and CO as predicted by DiscEvolution (Fig. \ref{fig:1Dcode_abu}) are taken as the initial conditions for the second step of the code: the chemistry and thermal balance. At each radius, the output abundance from DiscEvolution is taken, with this same abundance being used for all cells in the vertical direction. The vertical mixing in the disk is assumed to be very efficient and instantaneous. With these initial conditions, the chemistry is run time-dependently for 1000 years. This time was chosen as it is significantly shorter than the time between our DiscEvolution snapshots, meaning that the transport of dust and gas will not have significantly changed the composition or dust distribution of the disk in the meantime. Additionally, it is long enough for the chemistry in the surface layers to have reached steady-state. {Test calculations have shown that this is long enough for the warm, observable surface layers: the main formation reactions of \ce{H2O} (discussed in Sect. \ref{subsec:res_temp}) have timescales of only a few years in this layer of the disk, and the photodissociation timescale is equally short in regions where \ce{H2O} cannot self-shield against the UV radiation \citep[see][]{vlasblom2025_H2O}.} For the remaining atomic species in our chemical network, as well as for the grids in which the abundances are not taken from DiscEvolution, the chemistry for each model is instead started with all species (including hydrogen) in their atomic forms, with an abundance that is constant throughout the disk \citep[see][]{bosman2022b, vlasblom2024}, after which the chemistry is run for $10^3$ yr.

The chemical network used in this work is from \citet{miotello2016}, a network based on UMIST06 \citep{woodall2007} that includes isotope chemistry through the implementation of isotope-selective photodissociation and fractionation reactions \citep[see][]{miotello2014}. We amend this network with the changes made in \citet{bosman2022a, bosman2022b}, which include the effects of \ce{H2O} UV shielding and more efficient \ce{H2} formation at high temperatures through the inclusion of several three-body reactions. To generate synthetic spectra of \ce{H2O}, \ce{^12CO2}, and \ce{^13CO2}, DALI's ``fast ray-tracer'' \citep[see][Appendix B]{bosman2017} is used, assuming a distance of 121 pc and a face-on orientation. The excitation of all molecules was calculated in non-LTE. The molecular data file for \ce{H2O} is retrieved from the Leiden Atomic and Molecular Database (LAMDA), including levels with energies up to 7200 K \citep{tennyson2001}. The line transitions are obtained from the BT2 list \citep{barber2006} and the collisional rate coefficients come from \citet{faure2008}. For \ce{CO2} and its isotopologues, we use the molecular data compiled by \citet{bosman2017}, who retrieve the energy levels, line positions and line strengths from the HITRAN database \citep{rothman2013}. For \ce{^12CO2} and \ce{^13CO2}, collisional rate coefficients based on \citet{allen1980}, \citet{nevdakh2003}, and \citet{jacobs1975} are used. The generated spectra are convolved to the approximate resolving power of MIRI at $R=3000$, and no synthetic noise is added. 


\begin{figure*}
    \centering
    \includegraphics[width=0.9\linewidth]{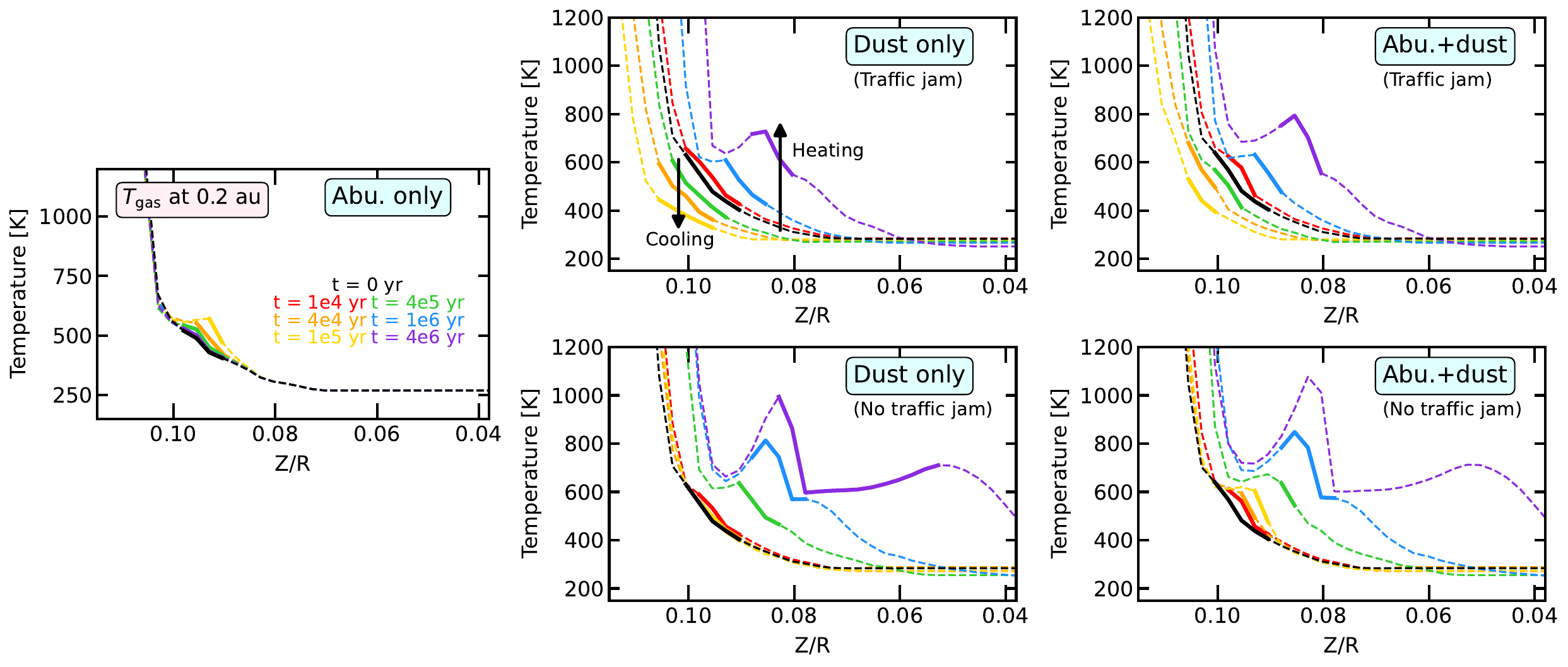}
    \caption{Gas temperature as a function of disk height in a vertical cut at 0.2 au for all grids at all time steps. The location of the \ce{H2O} emitting region is indicated by the {solid} parts of the lines.}
    \label{fig:tgas}
\end{figure*}

\subsection{Slab retrievals}\label{subsec:methods_slabs}

Finally, we follow the methods presented in \citet{vlasblom2025_H2O} and perform retrievals with LTE slab models on our synthetic \ce{H2O} and \ce{CO2} spectra. We refer the reader to that work for the full details. In summary, the slab models contain three free parameters: the molecular excitation temperature $T_{\rm ex}$ (usually close to the kinetic gas temperature $T$), the column density $N$, and the emitting area $A$ \citep[e.g.][]{tabone2023, kamp2023}. Assuming $A = \pi R_{\rm eq}^2$, the emitting area can be converted to an equivalent slab radius {($R_{\rm eq}$)}. This may correspond to the true emitting radius of the emission, but the emission can also originate from an annulus further out in the disk instead. 

The emission of \ce{CO2} and its isotopologues are fit with a single model, as their emission features span a narrow wavelength range. The best-fit model was obtained from a grid of slab models using a $\chi^2$ method. 

\ce{H2O}, on the other hand, has emission features spanning over a broad range in wavelength and upper level energy $E_{\rm up}$, and therefore requires a slightly more sophisticated setup. Following \citet{temmink2024b}, we fit the entire \ce{H2O} spectrum from 10-28 $\mu$m using three components, where the flux from three different slab models with decreasing temperature and increasing emitting area is combined, and the best fit is found using the Monte Carlo Markov Chain (MCMC) implementation emcee \citep{foreman_2013_emcee}. We select our fitting windows centered on several isolated lines across the wavelength range, which were identified by \citet{banzatti2025} in their Tables 5 and 6. The effects of mutual shielding of adjacent lines \citep[as done in, e.g.,][]{tabone2023} are not accounted for in any of our fits, as this is also not accounted for when generating our synthetic DALI spectra.

\section{Results}\label{sec:results}

\subsection{Gas temperature structure}\label{subsec:res_temp}

Fig. \ref{fig:tgas} presents the gas temperature in a vertical cut through the disk at 0.2 au (a radius right within the emitting region of warm \ce{H2O}) for all 5 grids of models at all time steps. {Fig. \ref{fig:tgas_map} presents the full 2D gas temperature maps for the two grids with varied abundances and dust (the two rightmost panels of Fig. \ref{fig:tgas}).} Considering first the grid in which only the abundances are varied (leftmost panel {of Fig. \ref{fig:tgas}}), it is clear that the temperature remains relatively constant throughout the different time steps, with the exception of a small temperature increase around $Z/R$ = 0.09-0.10. This layer in the disk is precisely co-located with the IR emitting area of \ce{H2O} (which is indicated in Fig. \ref{fig:tgas} by the bold regions of the lines). This layer of the disk has been shown to be particularly sensitive to changes in the thermochemistry \citep[see also][]{woitke2018, bosman2022b, vlasblom2024}. Namely, it represents the transition from the hot, atomic layer where the gas and dust temperatures are strongly decoupled into the deeper molecular layer where the gas and dust become coupled.

\begin{figure*}
    \centering
    \includegraphics[width=0.85\linewidth]{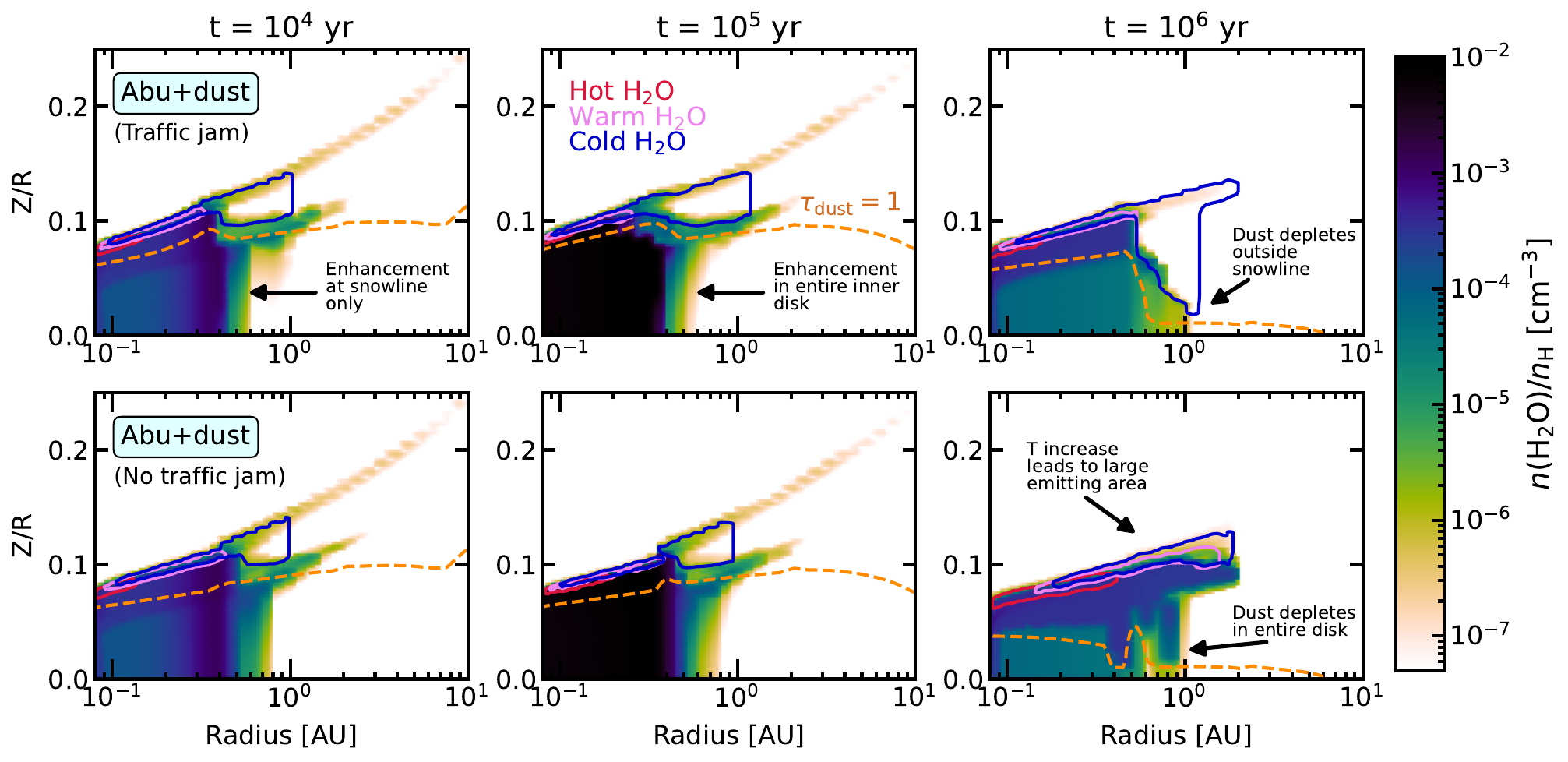}
    \caption{\ce{H2O} abundance as a function of radius and height at 3 time steps for the two grids with varied abundances and dust. In all panels, the dust $\tau=1$ surface at 15 $\mu$m is indicated with an orange dashed line. The red, pink, and blue contours in all panels represent the 80\% emitting regions of the \ce{H2O} 16$_{8,9}$ -- 15$_{7,8}$ line ($E_{\rm up}$ = 6052 K at $\lambda=17.324\;\mu$m), 13$_{4,9}$ -- 12$_{3,10}$ line ($E_{\rm up}$ = 3646 K at $\lambda=17.504\;\mu$m), and 8$_{3,6}$ -- 7$_{0,7}$ line ($E_{\rm up}$ = 1447 K at $\lambda=23.817\;\mu$m) lines, respectively, representing emission from hot, warm, and cold \ce{H2O}.}
    \label{fig:H2O_abu}
\end{figure*}

In the surface of the disk, the oxygen content of the disk exists primarily in its atomic form due to the photodissociation of \ce{H2O}. When moving down towards the midplane, this oxygen is gradually transformed from its atomic form in the upper layers first into OH via the \ce{O + H2 -> OH + H} reaction, and then into \ce{H2O} via the \ce{OH + H2 -> H2O + H} reaction. Around 0.1 Myr in our DiscEvolution simulations, the influx of \ce{H2O}, and thereby the influx of oxygen, into the inner regions of the disk has reached its peak (see yellow line in Fig. \ref{fig:1Dcode_abu}). As a result, the \ce{O + H2 -> OH + H} and \ce{OH + H2 -> H2O + H} consume much more of the \ce{H2} in this transition layer than at other time steps, replacing it with atomic H and thereby shifting the H/\ce{H2} transition deeper into the disk. At the same time, this extra atomic H can still reform into \ce{H2} via its formation route on the dust grains, which releases heat into the gas \citep{cazaux2002, cazaux2004}. These effects combined lead to the small temperature increase that is observed at 0.1 Myr. 

By contrast, however, the temperature changes are much larger in the other four grids in which the dust properties are varied between the different time steps. In the two grids where a traffic jam is present (Fig. \ref{fig:tgas}; top row, middle and right panels), the temperature is seen to first decrease (compare red and yellow lines) due to the build-up of dust in the inner regions of the disk caused by radial drift (see also Fig. \ref{fig:ndust}). As $T_{\rm gas}>T_{\rm dust}$ in this layer, this addition of dust leads to an increase in the thermal accommodation between the dust and the gas, therefore lowering the gas temperature. At later times, the temperature increases again (compare green and blue lines) due to the removal of dust from the disk by accretion onto the star. In the two grids without a traffic jam (bottom row, middle and right panels), this initial build-up of dust does not occur, and therefore the temperature does not change much at early times either, since not much dust has been removed by drift yet. At late times, once the outer disk runs out of pebbles with which to resupply the inner disk, the lack of a traffic jam in the inner disk leads to very efficient removal of dust from the disk (perhaps unrealistically so, since no dust traps are assumed to be present in our models). This causes the temperature to increase very strongly beyond 1 Myr in these models. 

Comparing the grids in which only the dust is varied to the grids in which both the dust and the abundances are varied, it is clear that the dust properties of the disk have a much larger effect on the gas temperature structure than the chemistry does. Still, the effects do combine, as seen by the small temperature increase seen at early times (0.1 Myr) in the bottom right panel of Fig. \ref{fig:tgas}. These effects also clearly change the depth and temperature from which the \ce{H2O} emits in the disk. When a dust pile-up cools the surface layers of the disk, the \ce{H2O} emits from further up in the disk, and the reverse is true when the surface layers are heated by a depletion of dust. We note that the \ce{H2O} emitting region in the 4 Myr model (purple line) of the bottom-right panel is not indicated in the figure. This is due to the strong increase in temperature leading to the \ce{H2O} emitting region moving to larger radii than 0.2 au.

\subsection{\ce{H2O} abundance, flux, and retrievals}\label{sub:h2o_abu}

Fig. \ref{fig:H2O_abu} presents the gas-phase \ce{H2O} abundance throughout the disk in two of our five grids (the two grids with varied dust and abundances), shown at three time steps. The full version of this figure with all five grids is presented in Fig. \ref{fig:H2O_abu_full}. At 10 kyr in the DiscEvolution simulations, the \ce{H2O} abundance has started building up just inside the midplane snowline (see red line in Fig. \ref{fig:1Dcode_abu}), and this can be seen as well at the midplane in the 2D abundance maps (see first column of Fig. \ref{fig:H2O_abu}). At 0.1 Myr, the \ce{H2O} abundance has reached its maximum at $\sim10^{-2}$ with respect to hydrogen, an enhancement of roughly a factor 100 compared to its initial abundance of $\sim10^{-4}$. At 1 Myr, the \ce{H2O} abundance has decreased again due to the viscous advection of the enhanced gas onto the star. 


\begin{figure*}
    \centering
    \includegraphics[width=\linewidth]{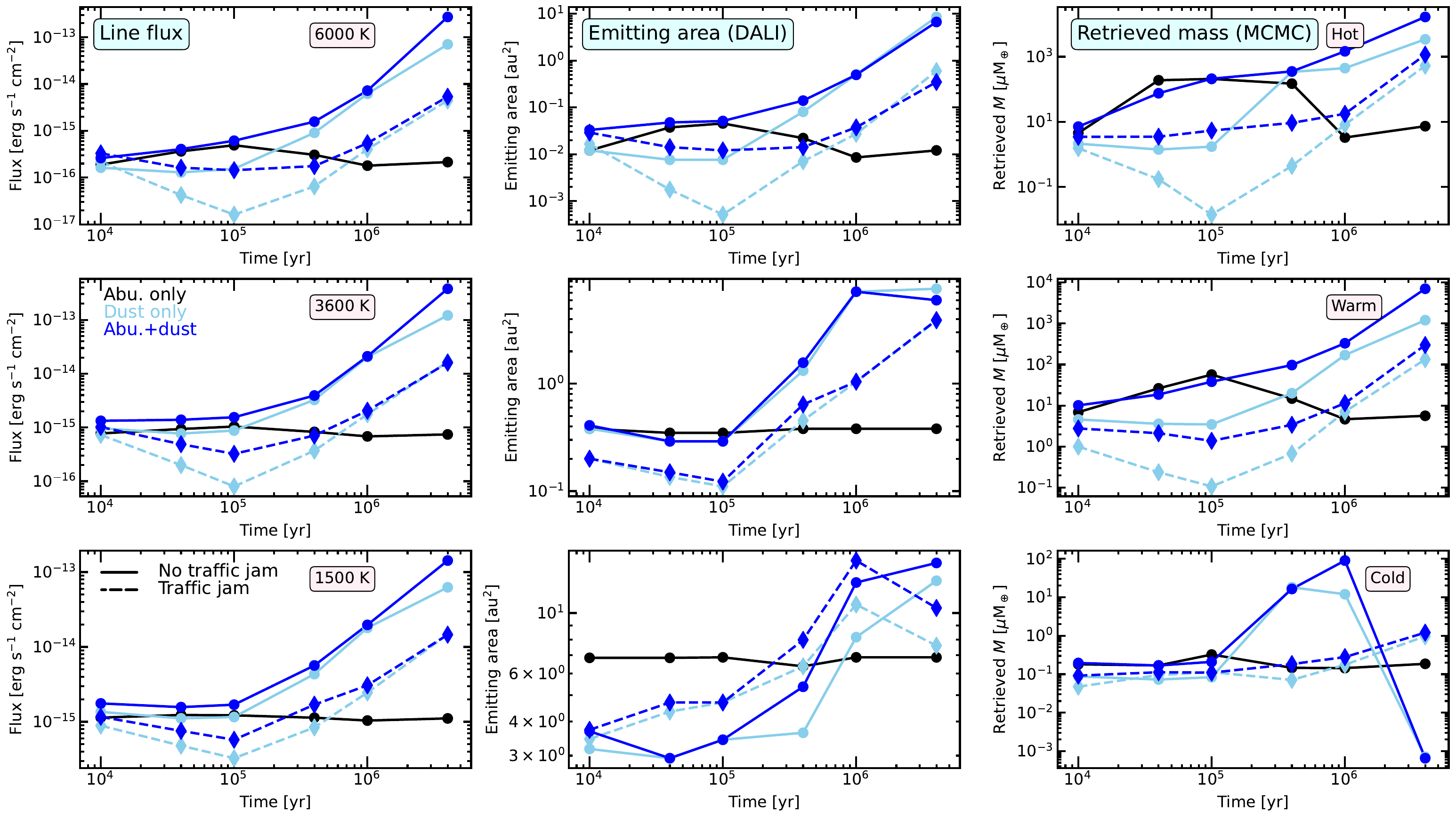}
    \caption{Left: Integrated \ce{H2O} flux of the 16$_{8,9}$ -- 15$_{7,8}$ line ($E_{\rm up}$ = 6052 K at $\lambda=17.324\;\mu$m; top panel), the 13$_{4,9}$ -- 12$_{3,10}$ line ($E_{\rm up}$ = 3646 K at $\lambda=17.504\;\mu$m; middle panel), and the 8$_{3,6}$ -- 7$_{0,7}$ line ($E_{\rm up}$ = 1447 K at $\lambda=23.817\;\mu$m; bottom panel) as a function of time for all grids. Middle: Emitting area of the three lines as determined from the DALI models. Right: Retrieved mass for each temperature component of a three-component MCMC fit, calculated as $M = N \times \pi R^2 \times m_{\rm \ce{H2O}}$. In each panel, the black {markers} represent the grid with varied abundances only, the light blue {markers} represent the two grids with varied dust only, and the dark blue {markers} represent the two grids with both dust and abundances varied. The {diamonds} indicate the grids in which a traffic jam is present{, and the dots represent those in which no traffic jam is present. Connecting lines have been added to guide the reader's eye}. Note the different vertical scales along each column.}
    \label{fig:H2O_retrievedN}
\end{figure*}

The dust distribution also has an impact on the \ce{H2O} abundance structure, mainly leading to (subtle) changes in the vertical extent, and therefore emitting height (see bold regions in Fig. \ref{fig:tgas}) of the gas-phase \ce{H2O}. At late times ($>$1 Myr), the depletion of dust leads to more significant changes, as the disk heats up significantly. In the models with a traffic jam, this is less prominent as a significant amount of dust still remains within the \ce{H2O} snowline. However, in the models without a traffic jam, the depletion of dust leads to the gas-phase \ce{H2O} reservoir, and therefore also the emitting area, extending to larger radii ($>$1 au).

The left column of Fig. \ref{fig:H2O_retrievedN} demonstrates how the \ce{H2O} line flux changes as a function of time in all five of our grids. We consider here the integrated flux of three lines identified in \citet{banzatti2025} to trace different temperature regimes: the 16$_{8,9}$ -- 15$_{7,8}$ line ($E_{\rm up}$ = 6052 K; tracing hot $\sim800$ K emission), the 13$_{4,9}$ -- 12$_{3,10}$ line ($E_{\rm up}$ = 3646 K; tracing warm $\sim400$ K emission), and the 8$_{3,6}$ -- 7$_{0,7}$ line ($E_{\rm up}$ = 1447 K; tracing cold $\sim200$ K emission). Generally, all three lines show a very similar trend with time. 

In the grid with varied abundances (black lines in Fig. \ref{fig:H2O_retrievedN}), the \ce{H2O} flux of all three lines shows a small increase around 0.1 Myr. This is caused in part by the increase in \ce{H2O} abundance but appears to be mostly driven by the increase in emitting area of each line (due to the increased gas temperatures), as shown in the middle column of Fig. \ref{fig:H2O_retrievedN} (where the values are taken directly from the DALI models). This is also evident from the right column of Fig. \ref{fig:H2O_retrievedN}, where we present the retrieved mass of the three temperature components from our MCMC fits to the \ce{H2O} spectra, which is calculated as $M = N \times \pi R^2 \times m_{\rm \ce{H2O}}$. The results of the MCMC fits, the retrieved temperature, column density, and emitting radius, are shown in Fig. \ref{fig:H2O_retrievedT}. All three temperature components of the fit show an increase in mass around 0.1 Myr, with the effect being the most prominent in the component tracing the hottest \ce{H2O} reservoir. However, the increase in flux is quite small (less than a factor 3 for all three lines), despite the \ce{H2O} abundance increasing by roughly a factor 100 (see Fig. \ref{fig:1Dcode_abu}). This clearly indicates that the three lines considered here have become optically thick, and therefore do not trace the full underlying enhancement in \ce{H2O}. Rather, it is likely that they mainly trace the increase in temperature that occurs around this time. 

In the two grids where only the dust properties are varied (light blue lines in Fig. \ref{fig:H2O_retrievedN}), we see a clear difference between the scenarios with and without a traffic jam (dashed and solid lines, respectively). When a traffic jam is present, a clear decrease in flux is seen in all three lines. This is matched by the retrieved mass of the hot and warm components, which show a clear decrease due to the enhanced optical depth caused by the build-up of dust. This is again more prominently seen in the 6000 K line, as well as the hottest component of the MCMC, as the dust pile-up is most strongly concentrated in its closer-in emitting region. The coldest component of the MCMC, however, behaves slightly differently to the flux of the 1447 K line. The retrieved mass of the coldest component is relatively constant, likely due to this component tracing further out into the disk, beyond the midplane \ce{H2O} snowline. This causes it to be less affected by the pile-up of dust. The 1447 K line, however, still shows a dip in its flux. This indicates that its flux is not entirely dominated by the coldest component of the MCMC, but rather it represents a combination of the warm and cold components. 

In the grid without the traffic jam, the pile-up of dust is not present and therefore the \ce{H2O} flux remains relatively constant at early times, until it strongly increases at later times due to the strong increase in temperature, increase in emitting area, and decrease in optical depth caused by dust depletion. This is consistent with the results of \citet{greenwood2019_dustevol}, and the retrieved mass follow the same trend as well. The two grids in which both the abundances and the dust properties are varied generally follow the trend of their dust-only counterparts closely, as the effects of the dust properties seem more significant in setting the \ce{H2O} flux than the chemistry, due to the lines rapidly becoming optically thick when the \ce{H2O} abundance is increased.

\begin{figure*}
    \centering
    \includegraphics[width=0.9\linewidth]{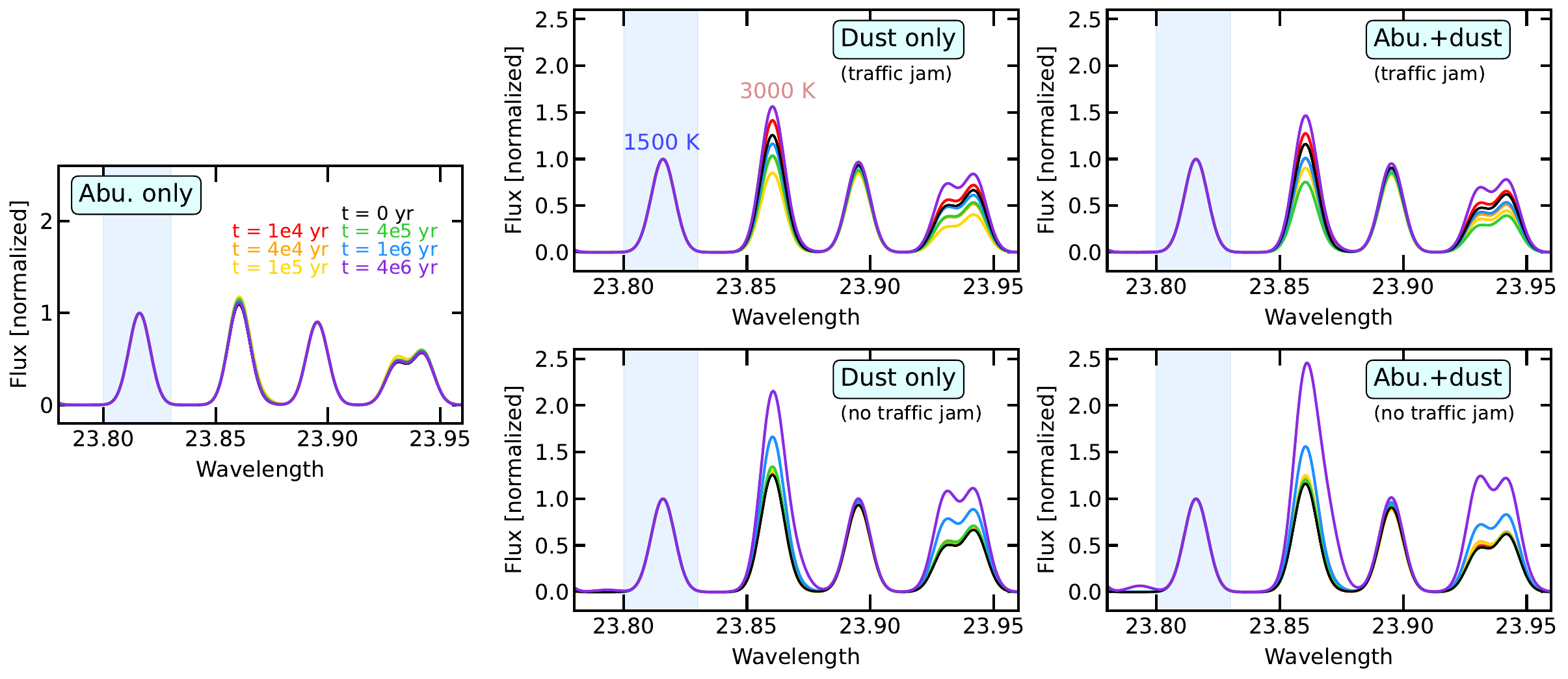}
    \caption{Synthetic \ce{H2O} spectra between 23.75 and 23.95 $\mu$m for all grids at all time steps. The flux is normalized to the 8$_{3,6}$ -- 7$_{0,7}$ line ($E_{\rm up}$ = 1447 K) at 23.82 $\mu$m, shaded in light blue. {Note that the absolute fluxes of these lines changes by up to two orders of magnitude with time (see bottom-left panel of Fig. \ref{fig:H2O_retrievedN}).}}
    \label{fig:H2O_23um}
\end{figure*}

Overall, the retrieved masses of the hot and warm \ce{H2O} components present a clear story. When only the abundances are varied, the retrieved mass clearly traces the influx of \ce{H2O} into the inner disk at early times and its subsequent draining onto the central star at later times (black lines). When the dust properties are varied, the build-up of dust around 0.1 Myr when a traffic jam is present leads instead to a decrease in retrieved mass (light blue dashed lines). On the other hand, when no traffic jam is present, the retrieved mass instead remains relatively constant at early times, until it increases strongly at later times when the dust depletes from the disk and more of the gas-phase \ce{H2O} reservoir is exposed (light blue solid lines). When both the dust and the abundances are varied simultaneously, we see that these effects combine. In particular, with a traffic jam present (dark blue dashed lines), the retrieved mass remains roughly constant until it increases at late times ($>$1 Myr), demonstrating how the effects of the dust pile-up and the influx of gas-phase \ce{H2O} cancel each other out almost perfectly. This is thus very consistent with the predictions made in \citet{sellek2025_CO2} and \citet{houge2025_H2O}, where the \ce{H2O} is indeed `smuggled unnoticed' into the inner disk. However, without a traffic jam present (dark blue solid lines), the delivery of \ce{H2O} into the inner disk is clearly visible in the retrieved mass. 

It is clear that the differences in retrieved mass are more pronounced in the hot component than the warm component, once again indicating that this reservoir is more sensitive to the changes in disk properties. The coldest \ce{H2O} component seems to portray a different behavior, which can likely be attributed to it tracing further out in the disk. In most of our grids, the retrieved cold \ce{H2O} mass remains relatively constant. Its column density is also relatively low, as this component is generally found to be the most optically thin of the three, which is in line with findings of observational works \citep[see, e.g.,][]{temmink2024b}. In the two grids with varied dust properties without a traffic jam, the mass of the cold component increases at late times. This is potentially due to the depletion of dust from the disk exposing more of the colder, deeper-lying reservoir. Then, at 4 Myr, the strong increase in temperature causes the emission from this reservoir to be drowned out again by emission from warmer components.

Whereas the retrieved \ce{H2O} mass (and therefore the column density) tells a clear story, the retrieved gas temperatures lack a clear trend. Fig. \ref{fig:H2O_retrievedT} demonstrates that the retrieved temperature remains roughly constant with time for most grids, and may even display behaviors that contradict the physical picture described above. For example, the two grids with a traffic jam display a peak in temperature at 0.1 Myr in the hot \ce{H2O} component, when the pile-up of dust is at its peak and the gas temperature of the disk is shown to decrease. The lack of a clear trend in retrieved temperature likely indicates that these fits may not provide the best indicators of how the different \ce{H2O} temperature reservoirs behave with respect to one another. Instead, line ratios likely provide a better indication of the relative strength of these reservoirs.

\subsection{Cold \ce{H2O} diagnostics}\label{sub:h2o_cold}

As the delivery of ices to the inner disk has been often proposed to be observable through tracers of the cold \ce{H2O} reservoir, it is interesting to consider how these behave in our model grids. In Fig. \ref{fig:H2O_23um}, four lines of interest between 23.7 and 24 $\mu$m are shown, which are a key diagnostic for cold \ce{H2O}. Two of these lines (the second and fourth lines) have an $E_{\rm up}$ of $\sim$3000 K, whereas the other two lines (the first and third lines) have a lower $E_{\rm up}$ of $\sim$1500 K. The relative strength of these four lines is very sensitive to the temperature of the gas \citep[e.g.,][]{banzatti2023, banzatti2025, temmink2024b}. When the cold, $\sim$200 K \ce{H2O} reservoir contributes strongly to the total flux (e.g. due to a large influx of sublimated \ce{H2O} ices near the snowline), the 1500 K lines are expected to be bright compared to the 3000 K lines. 

\begin{figure*}
    \centering
    \includegraphics[width=0.9\linewidth]{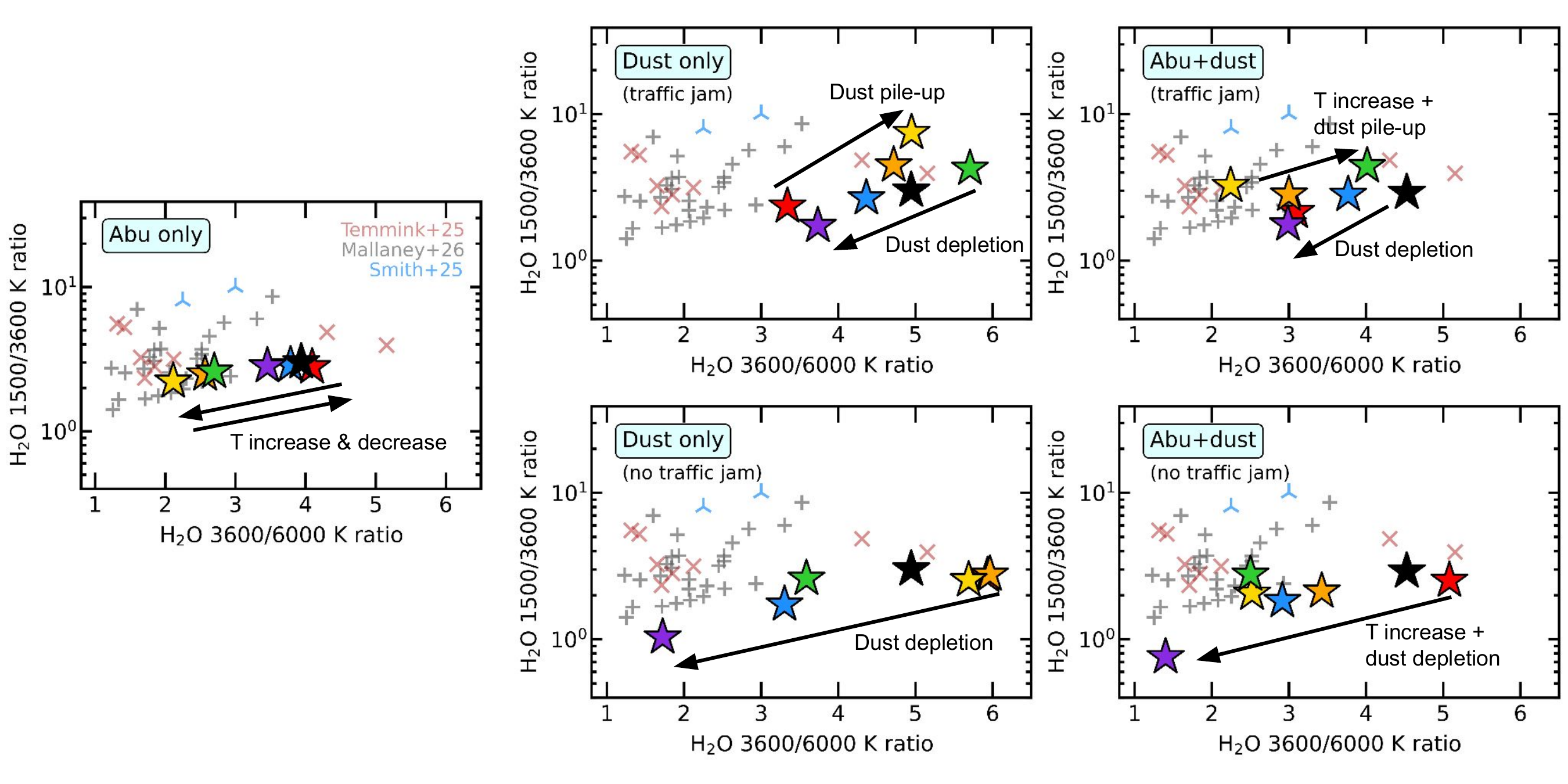}
    \caption{\ce{H2O} temperature diagnostic diagram as defined in \citet{banzatti2025} for all grids. The colored stars represent the different time steps in our grids, with the colors matching those shown in Fig. \ref{fig:tgas} (red = early, blue = late). The observations presented by \citet{mallaney2026}, \citet{temmink2025}, and \citet{smith2025} are shown as gray plus symbols, red crosses, and blue triangles, respectively.}
    \label{fig:H2O_B25}
\end{figure*}

Fig. \ref{fig:H2O_23um} presents the relative strength of this quartet of lines, normalized to the flux of the 8$_{3,6}$ -- 7$_{0,7}$ line ($E_{\rm up}$ = 1447 K; blue shaded region). Interestingly, the grid in which only the abundances are varied (and which should therefore most clearly show the effects of \ce{H2O} delivery by pebble drift) does not show any clear variation in this cold \ce{H2O} line ratio (leftmost panel). {The difference in line ratio is on the order of $\lesssim$10\%, which corresponds to the approximate precision with which individual line fluxes can usually be determined in JWST-MIRI observations at these wavelengths \citep[see, e.g.,][]{banzatti2025, arulanantham2025}. Therefore, these variations would not be detectable above typical noise levels.}

{The lack of variation} is not surprising, however, as \citet{vlasblom2025_H2O} demonstrated that the strength of the 1500 K lines is primarily sensitive to a \ce{H2O} reservoir further out in the disk ($>$ 1 au), above the snow surface. In our models, however, the midplane \ce{H2O} snowline is located at $\sim$0.5 au, and any enhancement in \ce{H2O} by drift is only mixed up vertically, thereby staying within 0.5 au. Therefore, the \ce{H2O} abundance in this reservoir traced by the cold \ce{H2O} lines does not change much in our models, and the relative strength of the 1500 and 3000 K lines also remains unchanged. This is also why the retrieved mass of the cold component showed the smallest changes with time (see Fig. \ref{fig:H2O_retrievedN}). Whereas this is caused by our current modeling assumptions, it also clearly demonstrates what is needed for the \ce{H2O} delivered by drift to truly be observed with cold \ce{H2O} tracers with JWST: the gas needs to be mixed upwards vertically, after which it must also diffuse outward radially. This demonstrates that a more detailed combined treatment of vertical and radial mixing, both inwards and outwards, may be important to include in future work. 

The grids with varied dust properties, however, do demonstrate a clear change in the line ratio {which is $\gtrsim$50\% (and would therefore be readily detectable)}. When a traffic jam is present, the relative strength of the 3000 K lines decreases significantly as the dust pile-up reaches its peak at 0.1 Myr (top panel of the middle column, yellow line). This is caused by the preferential obscuration of these lines by dust, as they trace mainly the warm \ce{H2O} reservoir which is located further in. As the 1500 K lines trace a larger emitting area, they are less affected by the pile-up of dust which is only located interior to the midplane \ce{H2O} snowline. Hence, an increase in relative strength of the 1500 K lines can be attributed solely to a delivery of dust, rather than a delivery of \ce{H2O}, in this case. 

Correspondingly, in the scenario where no traffic jam is present, this effect is not observed. The line ratio remains relatively constant until at late times ($>$1 Myr), the 3000 K lines increase in strength due to the increase in disk temperature. When both the dust and the abundances are varied simultaneously (right column of Fig. \ref{fig:H2O_23um}), the line ratio follows a similar behavior to its dust-only counterpart, again indicating that the effects from the dust are most important in setting this line ratio.

Fig. \ref{fig:H2O_B25} presents a temperature diagnostic diagram following \citet{banzatti2025}, using the three lines described above (as well as the 8$_{4,5}$ -- 7$_{1,6}$ line, with $E_{\rm up}$ = 1615 K). Here, the 3600/6000 K line ratio is indicative of the relative strength of the warm, $\sim$400 K \ce{H2O} reservoir, and the 1500/3600 K line ratio represents the relative strength of the cold, $\sim$200 K reservoir. For reference, several JWST measurements of these ratios have been included as well \citep{mallaney2026, temmink2025, smith2025}. 

In the grid with varied abundances only, the change in the line ratios is mostly limited to the 3600/6000 K ratio. The 1500/3600 K line ratio, and therefore the relative strength of the cold \ce{H2O} reservoir, remains roughly constant with time. This is to be expected given the lack of a visible change in the cold lines at long wavelengths (Fig. \ref{fig:H2O_23um}). The hot and warm reservoirs, however, do appear to change in relative strength, with a decrease in the 3600/6000 K line ratio indicating that the hot reservoir is becoming most prominent at $\sim$0.1 Myr. This coincides with the peak of the \ce{H2O} influx into the inner disk, and the subsequent increase in inner disk temperature is likely what causes this change. Hence, the delivery of \ce{H2O} into the inner disk is being traced by the hot and warm \ce{H2O} lines in this case, rather than the cold lines.

When the dust properties of the disk are varied, a stark difference between the scenarios with and without a traffic jam is once again observed. When a traffic jam is present, both line ratios increase at early times as the dust piles up, and decrease once the dust drains from the disk, with the data points thus moving along a diagonal line up and down in the diagram. In \citet{vlasblom2025_H2O}, a similar trend was identified and linked to an increasing or decreasing gas/dust ratio. When the dust piles up at early times, the gas/dust ratio decreases strongly in the inner disk. This most strongly affects the hottest lines that trace the innermost regions, and affects the cold lines that trace furthest out in the disk the least. Hence, both the 3600/6000 K and 1500/3600 K line ratios increase, and then decrease again when the dust clears from the inner disk and the gas/dust ratio increases again. When no traffic jam is present, however, no initial decrease in gas/dust ratio occurs, and only the increasing gas/dust with the gradual clearing of dust from the disk remains. Therefore, both line ratios decrease over time.

This behavior becomes slightly more complicated once the combined effect of dust and gas transport is accounted for. The overall trends described above still hold globally, but the increase in temperature due to the influx of \ce{H2O} at 0.1 Myr also seems to have an effect in temporarily decreasing the 3600/6000 K ratio, making the hot reservoir more prominent than it would be if only a dust pile-up were (or were not) present. 

Finally, we also note that, when comparing these line ratios in our models directly to the JWST data in Fig. \ref{fig:H2O_B25}, our model data points generally lie slightly to the right compared to the data for most time steps. {These sources do not differ significantly from our models in, for example, spectral type or luminosity, so this is not likely to be the cause of this offset. Instead, }this potentially indicates a lack of hot \ce{H2O} emission in our models, which was also identified in the models by \citet{vlasblom2025_H2O}. {Differences in accretion properties may also play a role, as the accretion rate is known to correlate with the hot \ce{H2O} mass \citep{banzatti2023}. \citet{calahan2026} demonstrate that increasing the accretion rate in DALI models indeed yields an increased mass of hot water, which may bring the line ratios into better agreement.} Still, we stress that the trends observed between the models remain informative, even if there is some complexity that is not yet captured by our models. 

\section{Discussion}\label{sec:discussion}

\subsection{Cold \ce{H2O} emission as a tracer of the pebble mass flux}\label{subsec:disc_pebflux}

\begin{figure*}
    \centering
    \includegraphics[width=0.7\linewidth]{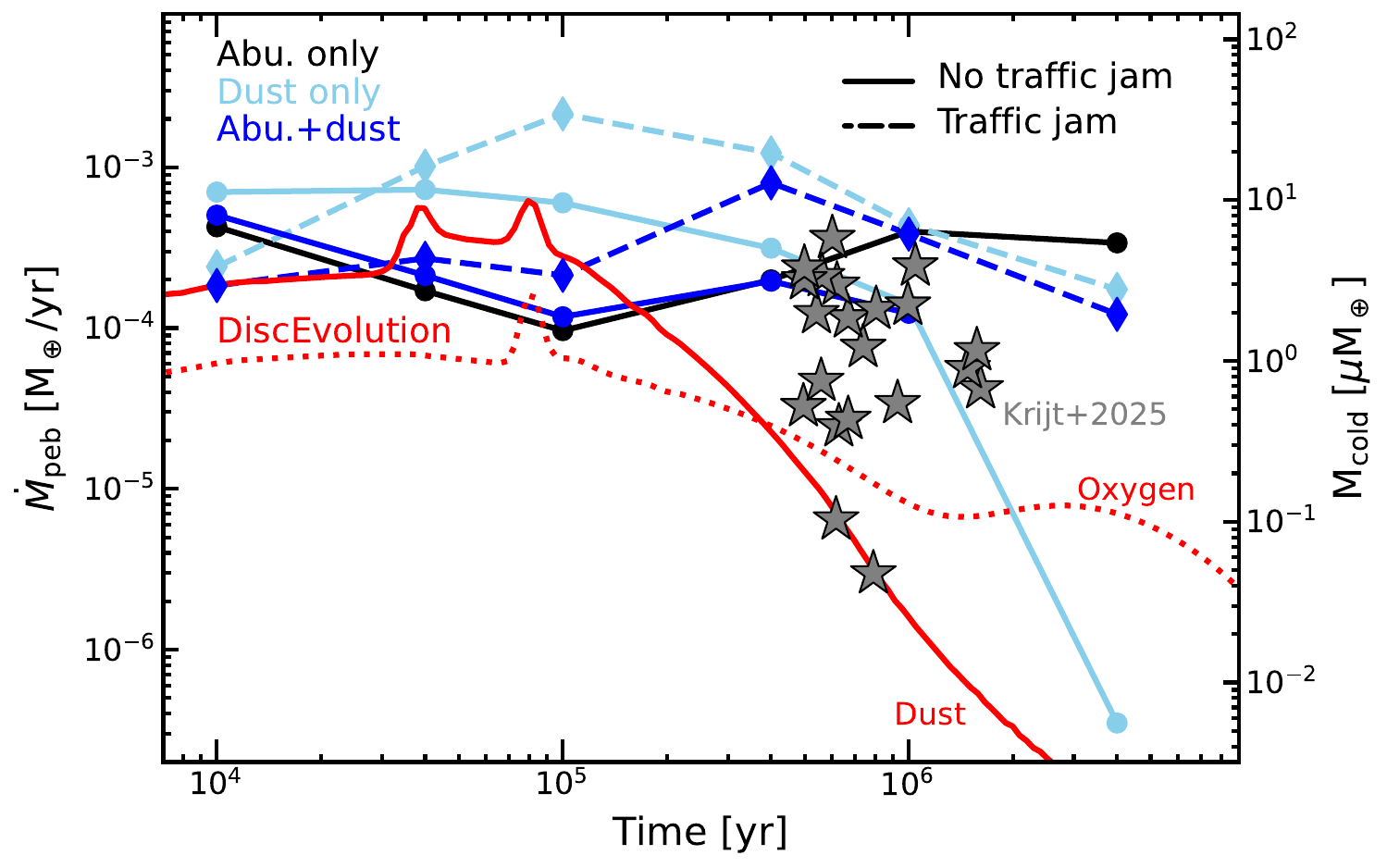}
    \caption{Pebble mass flux $\dot{M}_{\rm peb}$ as a function of time as derived from the 1500/6000 K line ratio (black, light and dark blue lines) and as calculated from the DiscEvolution simulations (red solid line). As a comparison, the oxygen mass flux (including both gas and ice) through the \ce{H2O} snowline as calculated from the DiscEvolution simulations is shown by the red dotted line. The gray {stars} represent the pebble mass fluxes derived from JWST-MIRI observations, as presented in \citet{krijt2025}. The right y-axis presents the cold \ce{H2O} vapor mass, related to the left y-axis via Eq. \ref{eq:Mpeb}. We note that, for the abu.+dust model without a traffic jam (dark blue solid line) at 4 Myr, no value of $M_{\rm \ce{H2O}}^{\rm cold}$ (and therefore $\dot{M}_{\rm peb}$) can be derived (see text).}
    \label{fig:H2O_Mpeb}
\end{figure*}

\subsubsection{General considerations}

Cold \ce{H2O} emission is often associated with the delivery of \ce{H2O} to the inner disk by pebble drift \citep[e.g.][]{romero-mirza2024_sample, banzatti2023, krijt2025}. However, this connection may not be entirely straightforward. When only the gas-phase abundances in the disk are varied, no significant enhancement in cold \ce{H2O} emission is seen concurrently with the delivery of gas-phase \ce{H2O} to the inner disk. After all, the reservoir from which the cold \ce{H2O} lines emit (a layer above the snow surface beyond 1 au, as identified in \citealt{vlasblom2025_H2O}) is not enhanced in abundance in our modeling approach, indicating that a more detailed treatment of vertical mixing including outward diffusion may be important for future work. Instead, the strength of the cold \ce{H2O} emission is most strongly impacted by the changes in dust properties when a traffic jam is present. The delivery of gas-phase \ce{H2O} is technically not necessary to produce a spectrum with strong cold \ce{H2O}, and it is the delivery of \textit{dust} rather than gas that is being traced by the cold \ce{H2O} lines in this case. 

Since dust likely carries \ce{H2O} ice, it is plausible to assume that the strength of the cold emission will still hold some relation to the efficiency of radial drift, and thereby the icy pebble flux reaching the inner disk. Recent work by \citet{romero-mirza2024_sample} and \citet{krijt2025} has proposed to link the strength of cold \ce{H2O} emission to a pebble flux crossing the \ce{H2O} snowline. In \citet{romero-mirza2024_sample}, the pebble flux $\dot{M}_{\rm peb}$ is linked to an observable cold \ce{H2O} vapor mass $M_{\rm \ce{H2O}}^{\rm cold}$ through the following equation:
\begin{align}
    \dot{M}_{\rm peb} = \frac{M_{\rm \ce{H2O}}^{\rm cold}}{t_{\rm \ce{H2O}}} \frac{\eta}{f_{\rm ice}f_{\rm \ce{H2O}}}. \label{eq:Mpeb}
\end{align}
This equation represents the equilibrium between the \ce{H2O} supplied by sublimation and that destroyed by photodissociation or other processes. $f_{\rm ice}$ is the mass fraction of pebbles that is made up by ices and $f_{\rm \ce{H2O}}$ is the mass fraction of said ices in \ce{H2O} ice. $\eta$ is a correction factor accounting for the portion of the total \ce{H2O} column located beneath an optically thick layer (which is unobservable). $t_{\rm \ce{H2O}}$ represents the approximate time that the delivered gas-phase \ce{H2O} will reside in the gas-phase, before it is removed or altered by processes such as diffusion, photodissociation, or chemical reactions. \citet{romero-mirza2024_sample} assume $f_{\rm ice} = 0.2$, $f_{\rm \ce{H2O}} = 0.8$, $\eta=10^3$, and $t_{\rm \ce{H2O}} = 100$ yr (though we note that these parameters, especially $\eta$, can have a rather large uncertainty). 

Thus, calculating the pebble flux from observational data hinges on the estimation of this cold \ce{H2O} vapor mass. \citet{krijt2025} present an empirical relation linking this quantity to the 1500/6000 K line ratio as defined by \citet{banzatti2025}, which takes the following form:
\begin{align}
    {\rm 1500/6000\;K} = \mathcal{A} + \left( \frac{M_{\rm \ce{H2O}}^{\rm cold}}{0.4\; \mu M_\oplus} \right)^{q}. \label{eq:Mcold}
\end{align}
\citet{krijt2025} assume $\mathcal{A}=1.75$ and $q=0.8$, which provided the best match with slab models. 

Our models provide us with access to both this 1500/6000 K line ratio from our DALI synthetic spectra and the true pebble flux crossing the \ce{H2O} snowline from our DiscEvolution simulations. Therefore, our models can obtain further insights into the link between the two quantities. This is demonstrated in Fig. \ref{fig:H2O_Mpeb}, where we present the pebble mass flux as derived from the 1500/6000 K line ratio (black, light blue, and dark blue lines), as well as the dust mass flux passing through the midplane \ce{H2O} snowline at 0.5 au (red solid line). 

Overall, the pebble flux estimates derived from this line ratio provide a decent order-of-magnitude approximation of the true pebble flux at early times. At later times, however, the 1500/6000 K line ratio overestimates the true pebble flux by several orders of magnitude in most grids. We do note that, for the two models with varied dust properties without a traffic jam (light and dark blue solid lines), at 4 Myr the 1500/6000 K line ratio either very closely approaches $\mathcal{A}=1.75$ (in the case of the dust-only model) or becomes smaller than this value (in the case of the dust+abundances model). In these cases, the derived cold \ce{H2O} mass and pebble flux will become very small or even unconstrained in the case of the dust+abundance model. We also note that there is no difference in pebble flux between the DiscEvolution simulations with and without a traffic jam. After all, we are measuring the dust mass flux passing through the \ce{H2O} snowline at 0.5 au, exterior to which the coupling of the gas and dust (and therefore the drift speeds) is kept exactly the same in both models. 

\subsubsection{Early times ($<$1 Myr)}

Most strikingly, Fig. \ref{fig:H2O_Mpeb} demonstrates that the estimation of the pebble mass flux from the 1500/6000 K line ratio can introduce complexities and trends as a function of time that do not match the evolution of the true DiscEvolution pebble mass flux. In particular, the dust pile-up in the dust-only model with the traffic jam (light blue dashed line) increases the 1500/6000 K line ratio significantly, thereby leading to a rise in the pebble mass flux that is much larger than what is present in the DiscEvolution models. From Fig. \ref{fig:H2O_retrievedN} we can see that this is caused by a combination of several effects. At this time, the temperature in the disk drops due to the build-up of dust. This causes the emitting area of the 6000 K line to shrink, whereas the emitting area of the 1500 K line slightly increases. Additionally, the added optical depth due to dust is also seen to only introduce a dip in the retrieved mass of the hot (and warm) component, not in the cold component of the MCMC. This indicates that the increase in line ratio, and therefore the perceived increase in pebble flux, is driven almost entirely by a loss of hot \ce{H2O} mass, rather than gained cold \ce{H2O} mass.

On the other hand, the model with varied abundances-only shows a dip in the derived pebble mass flux, exactly opposite to the true trend, which is once again caused by the increase in temperature and \ce{H2O} abundance that affects the 6000 K line the strongest. In this case, both the emitting area of the 6000 K line and the retrieved mass of the hot component are seen to increase (Fig. \ref{fig:H2O_retrievedN}), whereas for the 1500 K line and cold component these values remain constant. Hence, this dip in the retrieved pebble flux represents gained hot \ce{H2O} mass, rather than a loss of cold \ce{H2O} mass. Combining then the effects of the dust and the gas-phase abundances in the presence of a traffic jam (dark blue dashed line), the two effects somewhat cancel out. This is partially caused by the `smuggling unnoticed' of the incoming \ce{H2O}, as the enhanced \ce{H2O} column density and enhanced dust opacity cancel one another out, but it is also linked to the temperature changes in the disk. The increase in temperature due to the influx of oxygen (which leads to an increase of the emitting area of the 6000 K line) and the decrease in temperature due to the dust pile-up (leading to a decrease in the emitting area of the 6000 K line) also cancel each other out. Therefore, the overall agreement with the true pebble flux is improved at early times, but not for the right reasons. The changes in both optical depth and disk temperature caused by the transport of material, to which the 6000 K line is more sensitive than the 1500 K line, clearly bring additional complexities for the use of this line ratio as a proxy for the pebble mass flux. 

\subsubsection{Late times ($\geq$1 Myr)}

At late times ($\geq$1 Myr; the age range at which T Tauri disks are typically observed), almost all models predict a pebble flux that is much larger than the true value, by up to $\sim$3 orders of magnitude. Only the grids with varied dust without a traffic jam (light and dark blue solid lines) show a strong decrease in the 1500/6000 K line ratio at late times. While this is technically indeed reflective of the fact that these disks will have lost a lot of their dust mass due to drift at this point, the decreasing line ratio is once again mainly set by changes in the disk's optical depth and temperature. At late times, the decreasing dust optical depth leads to an increase in retrieved \ce{H2O} mass in both the cold and hot components. Consequently, the disk also heats up strongly, which leads the emitting area of both lines increasing sharply. However, this effect is much stronger for the 6000 K line, as its emitting area is seen to increase by two orders of magnitude, compared with an increase of only a factor few for the 1500 K line. Therefore, the line ratio decreases sharply.  

When comparing these derived pebble fluxes to those derived from JWST-MIRI observations (sample from \citealt{krijt2025} shown in gray diamonds in Fig. \ref{fig:H2O_Mpeb}), a similar phenomenon is observed. Most observed line ratios match our models more closely than the DiscEvolution pebble flux. Additionally, the disks which are found to have a lower predicted pebble flux are ones that host close-in gaps, which are expected to block a large part of the pebble flow. \citet{krijt2025} demonstrate that full disks, or disks with gaps located further out, generally have a larger 1500/6000 K line ratio (and thereby predicted pebble flux). Since our models do not include a gap, these disks are precisely the ones which make for the most fair comparison to our work. Hence, there seems to be an inherent mismatch between the predicted pebble flux from the 1500/6000 K line and the true pebble flux at late times. 

These findings are also consistent with recent results from \citet{xie2026}, who studied disks in the Upper Scorpius region (age $\sim$5 Myr) with JWST-MIRI. When comparing the 1500/6000 K line ratio between the older Upper Scorpius disks and younger disks (age $\sim$1-3 Myr) from the JDISCS survey, they find no meaningful difference. This is in line with our model predictions. Additionally, they note that two of their disks with low 1500/6000 K ratios are very \ce{H2O}-rich, and have high accretion luminosities. Thus, if their \ce{H2O}-rich spectra are indeed reflective of a high pebble flux, this lends further credence to the idea that changes in the hot \ce{H2O} mass may skew the line ratio towards lower values, thereby masking potential drift signatures from the cold reservoir.

\subsection{Implications for cold \ce{H2O} emission as a tracer of drift}\label{subsec:disc_impl}

The mismatch between the true and predicted pebble fluxes at late times warrants some further consideration, to evaluate whether any of the underlying equations (Eqs. \ref{eq:Mpeb} and \ref{eq:Mcold}) contain limitations that could be modified to account for further complexities revealed by this work. We consider here a few different effects that may be important to consider in future work.

First, Eq. \ref{eq:Mpeb} accounts for the effects of optical depth on the observable cold \ce{H2O} mass through the factor $\eta$. However, we have demonstrated that the dust optical depth of the disk changes a lot due to the changing dust distribution of the disk with time, and thus this factor is likely not constant with time. Naturally, the value of $\eta$ as a function of time then relates back directly to the pebble mass flux, as this will determine how fast a disk will lose its dust mass, or how long it may retain it. In fact, $\eta$ (and thus $\tau$) should depend positively on $\dot{M}_{\rm peb}$. This will weaken the dependence of $M_{\rm cold}$ on $\dot{M}_{\rm peb}$, which then also predicts the existence of a `worst case scenario' where there is no dependence of $M_{\rm cold}$ on $\dot{M}_{\rm peb}$ at all anymore, and the \ce{H2O} is smuggled unnoticed. Such a dependence may therefore be important to consider in future work. 

Second, Eq. \ref{eq:Mpeb} assumes all observed cold \ce{H2O} to be a product of sublimation. However, this neglects the effects of the gas-phase chemistry in the disk. The DiscEvolution models produce a strong depletion in gas-phase \ce{H2O} in the inner disk at the latest time considered in our grids, down to a fractional abundance of $\sim10^{-6}$ at 4 Myr (see Fig. \ref{fig:1Dcode_abu}, purple line in top left panel). However, the gas-phase \ce{H2O} abundance in the IR emitting layer of our corresponding DALI models does not deplete this far, instead remaining around a fractional abundance of $\sim10^{-4}$. This is caused by the fact that the gas-phase chemistry in the inner disk will rapidly produce \ce{H2O} as long as oxygen is present in some form. While the \ce{H2O} abundance in the inner disk decreases strongly at late times in the DiscEvolution models, CO gas from the outer disk replaces it. Therefore, the total amount of oxygen in the inner disk does not decrease as strongly. We demonstrate this in Fig. \ref{fig:H2O_Mpeb} by showing the mass flux of oxygen (gas and ice) through the \ce{H2O} snowline as a function of time (red dotted line). This oxygen mass flux is much more constant with time than the pebble mass flux, as it has only decreased by roughly one order of magnitude at 4 Myr, whereas the pebble mass flux has decreased by nearly three orders of magnitude. 

Therefore, some background level of cold \ce{H2O} can remain present in the inner disk due to gas-phase chemistry, even when the inflow from icy pebbles has ceased almost entirely. Fig. \ref{fig:H2O_abu} provides more evidence for this through the fact that the \ce{H2O} abundance is higher in the surface layers than in the midplane by 1 Myr. In the midplane, the chemical timescales are longer and therefore the conditions more closely reflect the initial conditions from the DiscEvolution models. However, in the surface layers, \ce{H2O} is actively and rapidly being formed from other O-bearing species. Therefore, in disks with a low pebble flux, Eq. \ref{eq:Mpeb} may overestimate the pebble flux. Instead, the observed cold water mass should be treated as an upper limit on the sublimated cold mass, and thus the derived pebble flux.


Finally, the analysis of \citet{krijt2025} demonstrate the 1500/6000 K line ratio to be sensitive not just to $M_{\rm \ce{H2O}}^{\rm cold}$, but also to $M_{\rm \ce{H2O}}^{\rm hot}$ (see Fig. A.1 in that work). At low values of $M_{\rm \ce{H2O}}^{\rm cold}$, the scatter is rather minimal, but it grows substantially when $M_{\rm \ce{H2O}}^{\rm cold}$ increases. This matches the findings of our work (see Sect. \ref{subsec:disc_pebflux}), as we have demonstrated that changes in the 1500/6000 K line ratio are often driven by changes in \textit{both} the hot and cold \ce{H2O} reservoirs, in some cases even more so by the hot reservoir than the cold. Still, Eq. \ref{eq:Mcold} is only calibrated against $M_{\rm \ce{H2O}}^{\rm cold}$. For future work, it may be important to take into account the changes induced by transport on $M_{\rm \ce{H2O}}^{\rm hot}$ as well.

To summarize, while the 1500/6000 K line ratio may provide a decent order-of-magnitude estimate of the pebble mass flux at early times, it can greatly overestimate the true pebble flux at late times. This line ratio is significantly influenced by the cold \ce{H2O} mass, but also the hot \ce{H2O} mass. The latter can change significantly due to changes in the disk's temperature and dust optical depth as a result of pebble drift. Hence, while the transport of icy grains can indeed affect this line ratio, directly linking it to the mass of pebbles crossing the snowline is not entirely straightforward. 

Naturally, one can try to circumvent this issue by not taking the 1500/6000 K line ratio as a proxy for the cold \ce{H2O} mass, and instead deriving it from slab models fits \citep[as done in][]{romero-mirza2024_sample}. This may indeed yield a better sensitivity to the total abundance as line optical depth (and thus saturation) plays less of a role (see Fig. \ref{fig:H2O_retrievedN}). Still, this does not guarantee an accurate derivation of the pebble flux from that mass. There may still be some dependence on $\eta$ or the disk chemistry that is not accounted for, which require further investigation to fully understand how best to derive a pebble flux from \ce{H2O} observations.

\begin{figure}
    \centering
    \includegraphics[width=\linewidth]{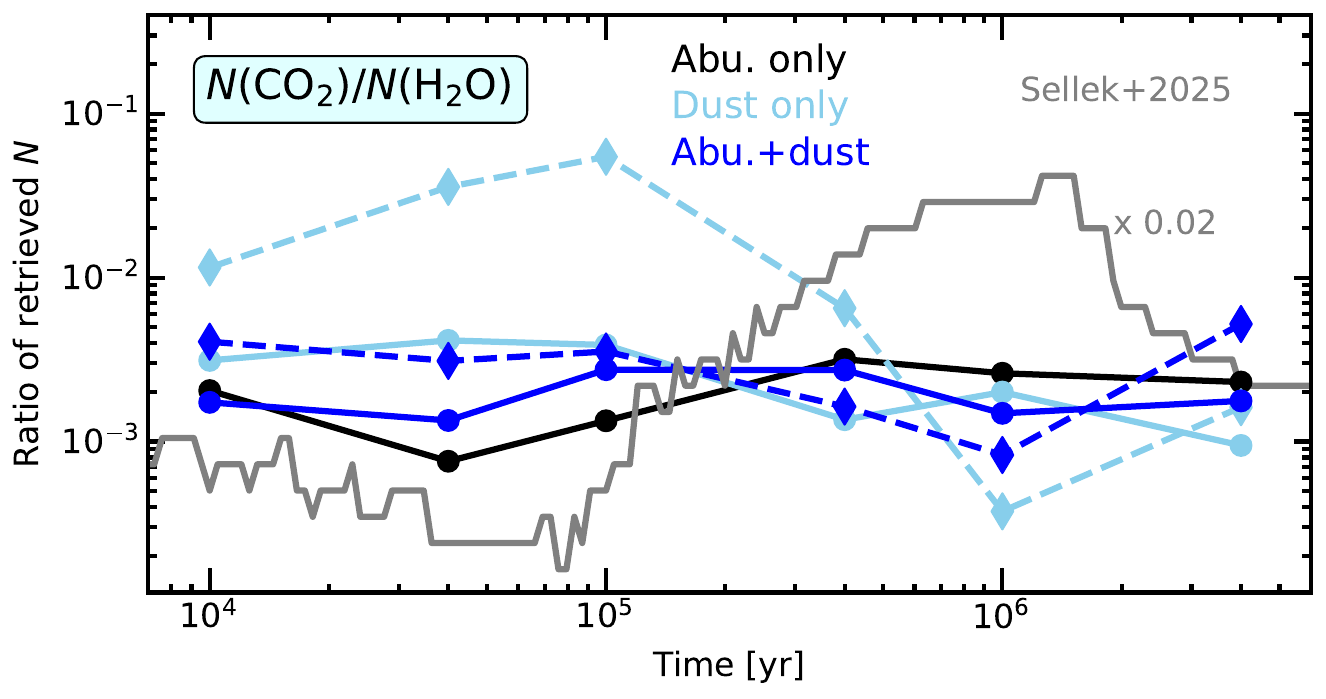}
    \caption{\ce{CO2}/\ce{H2O} retrieved column density ratio as a function of time for all grids. The {diamond markers with} dashed lines represent the models with a traffic jam, and the {round markers with} solid lines represent models without a traffic jam.}
    \label{fig:ratio_CO2only}
\end{figure}

\subsection{\ce{CO2}/\ce{H2O} as a tracer of drift}

The retrieved \ce{H2O} column can be strongly affected by the dust opacity in a traffic jam scenario, and become almost invariant with time. Therefore, \citet{sellek2025_CO2} propose an alternative diagnostic of \ce{H2O} delivery: the \ce{CO2}/\ce{H2O} column density ratio. Unlike the enhancement of \ce{H2O}, the enhancement of \ce{CO2} is not co-located with the traffic jam in space and time. Therefore, using this column density ratio may help address one of the issues highlighted in Sect. \ref{subsec:disc_impl}, the dependence of the pebble flux on $\eta$. Fig. \ref{fig:ratio_CO2only} presents the \ce{CO2}/\ce{H2O} retrieved column density ratio from our models. A full exploration of the \ce{CO2} abundance structure and the response of its spectral features to the disk evolution in each model can be found in Appendix \ref{app:CO2}. 

The results from \citet{sellek2025_CO2} are shown in the grey line in Fig. \ref{fig:ratio_CO2only}, which is a model that accounts for changes in both dust and gas and contains a traffic jam. Therefore, it should be compared to our dust+abundances model with a traffic jam (dashed dark blue line). Our model does not follow the proposed behavior in \citet{sellek2025_CO2}. This is caused by the abundance structure of \ce{CO2} being much more intricate than \ce{H2O} (see Fig. \ref{fig:CO2_abu}), and the most emissive \ce{CO2} is predominantly found in a thin layer where \ce{H2O} cannot yet self-shield. Therefore, the column density of \ce{CO2} is limited by the \ce{H2O} self-shielding, not the dust $\tau=1$ surface, and is therefore approximately constant as the dust properties are changed (see Appendix \ref{app:CO2}). 

However, this assumes a static disk without vertical or radial mixing, which is a caveat for future modeling. Additionally, several isotopologues of \ce{CO2} have been detected in a few disks \citep{grant2023, vlasblom2025_CXTau, frediani2025, salyk2025}, which can trace the deeper, colder reservoir further out (see Fig. \ref{fig:CO2_abu}), and therefore may be more promising for showing the proposed effect. They are also often found to be optically thin, so one may need to consider the total \ce{CO2} mass (or total number of molecules $\mathcal{N} = N\times\pi R^2$) rather than the column density, which is better constrained (see Appendix \ref{app:CO2}).

\subsection{{Further considerations: spectral types and mixed ices}}

{This work considers the evolution of only a single spectral type, as our models are based on the K5 star AS 209. However, modeling work has demonstrated that the evolution of the disk's composition may play out differently for stars of different spectral types \citep[e.g.][]{mah2023}. Depending on the spectral type of the central star, the spatial scales on which transport processes occur will be different as snowlines may be located closer in or further out. This, along with the mass of the star, will then also alter the timescales on which these processes occur. In particular, the lower stellar mass and closer-in snowlines of low-mass stars allow for the composition of the disk to evolve on much shorter timescales, meaning that they may become carbon-rich much sooner than T Tauri disks \citep[e.g.][]{mah2023, sellek2025_carbon}. }

{Still, the overall sequence plays out the same regardless: \ce{H2O} is brought in first, making the inner disk oxygen-rich, followed closely by \ce{CO2} and on longer timescales by carbon-rich gas, which raises the C/O ratio. Therefore, the trends observed in this work will likely remain, but may occur on shorter or longer timescales depending on the properties of the host star. On a population level, the presence of stars of different spectral types may make it more difficult to assess what stage of its evolution (\ce{H2O}-rich, \ce{CO2}-rich, carbon-rich) a population of disks is in, as it is likely that a range of conditions will be probed.}

{Another consideration that the spectral type of the star brings to mind is the strength of UV radiation, which in turn may alter the chemistry in the disk \citep[e.g.][]{walsh2015}. AS 209 is quite luminous in the FUV, with an FUV luminosity of $8\times10^{-3}$ L$_\odot$. \citet{vlasblom2025_H2O} tested the effects of both higher and lower FUV luminosity on the \ce{H2O} abundance in the disk, and found only a minor impact. For stars with a lower FUV luminosity, the strength of the cold \ce{H2O} lines may be slightly increased (by $\sim$20-30\%) due to cold \ce{H2O} being able to survive further out in the surface layers where it would otherwise be rapidly photodissociated. The hot \ce{H2O} reservoir may be slightly more affected in stronger FUV fields, as also discussed in Sect. \ref{sub:h2o_cold}. Still, this is unlikely to affect the trends observed in this work.}

{Finally, we have only considered the transport of pure ices in this work. However, \ce{H2O} ice also has the capability to trap some amount of volatiles such as CO, which will then desorb much closer to the star than they normally would \citep[see, e.g.,][]{collings2003, ligterink2024}. While the trapping of volatiles within \ce{H2O} ice will not affect the delivery of \ce{H2O} itself, it may affect the delivery of oxygen as a whole. In Fig. \ref{fig:H2O_Mpeb}, we demonstrate how the mass flow of oxygen is relatively constant with time, and therefore contributes to the continued presence of \ce{H2O} in the inner disk via gas-phase chemistry. If a large part of the volatile oxygen, for example in the form of CO, could be trapped in \ce{H2O} ice, it could potentially be delivered to the inner disk much sooner than it otherwise would. }

{However, only a limited amount of CO can be trapped by \ce{H2O} ice (roughly 5-20\% by mass; see \citealt{ligterink2024}). Additionally, transport models that include CO trapping by \ce{H2O} ice from \citet{williams2025} demonstrate that the gaseous CO released at the \ce{H2O} snowline will subsequently diffuse back outwards towards its own snowline, and therefore efficiently spread back out across almost the entire disk. The amount of gas-phase CO that advects onto the star at late times is thus not actually that significantly reduced. Therefore, the results from this work likely would not change significantly if trapping of CO ice were to be accounted for. }

\section{Conclusions} \label{sec:conclusions}

This work investigates how the emission of \ce{H2O} can be used as a tracer of pebble drift by combining the 1D transport code DiscEvolution with the 2D thermochemical code DALI. We summarize our main findings as follows:

\begin{itemize}
    \item The temperature structure of the disk is significantly affected by the transport of both gas and dust to the inner disk (Fig. \ref{fig:tgas}). The delivery of oxygen (mainly in the form of \ce{H2O}) to the inner regions brings the H/\ce{H2} transition deeper into the disk, and thereby raises the temperature in the IR emitting layers of the disk. A pile-up of dust in the inner disk, on the other hand, will lower the temperature in these layers. With time, as the dust depletes from the disk, the temperature in the entire disk rises strongly. 
    \item The IR emission from \ce{H2O} is affected by both the abundance and temperature structure of the disk (Fig. \ref{fig:H2O_retrievedN}, left column). However, as many IR lines are found to rapidly become optically thick as the abundance increases, trends in line flux generally follow the evolution of the disk's temperature structure. 
    \item The retrieved hot and warm \ce{H2O} mass from LTE slab fits shows effects from both the changes in abundance as well as the changes in dust properties of the disk (Fig. \ref{fig:H2O_retrievedN}, right column). When a traffic jam is present in the inner disk, the delivery of gas-phase \ce{H2O} can proceed mostly unnoticed due to the co-delivery of dust, as no change in retrieved mass is observed. When no traffic jam is present, the delivery can be traced through the change in retrieved mass. 
    \item The retrieved cold \ce{H2O} mass, on the other hand, is much more constant with time regardless of the dust properties of the disk, as it traces further out and is therefore less affected by a potential traffic jam.
    \item When only gas-phase \ce{H2O} is delivered to the inner disk, the relative strength of the cold \ce{H2O} lines at 23.8 $\mu$m (which are thought to trace pebble drift) does not change (Fig. \ref{fig:H2O_23um}, leftmost panel; see also Fig. \ref{fig:H2O_B25}). This indicates that \ce{H2O} sublimating at the snowline needs to undergo (rapid) radial outward diffusion combined with vertical mixing for observations of cold \ce{H2O} emission to truly trace \ce{H2O} that has been delivered to the inner disk.
    \item Instead, the relative strength of the cold lines is most strongly affected by changes in the dust distribution of the disk when a traffic jam is present (Fig. \ref{fig:H2O_23um}, top row; see also Fig. \ref{fig:H2O_B25}). The pile-up of dust in this scenario allows for the preferential obscuration of the hot and warm \ce{H2O} lines, thereby producing a spectrum with strong cold \ce{H2O} solely through the delivery of dust, rather than the delivery of \ce{H2O}.
    \item The use of the 1500/6000 K \ce{H2O} line ratio as a proxy for the pebble mass flux crossing the \ce{H2O} snowline warrants caution. It is found to provide a decent order-of-magnitude estimate of the pebble flux at early times ($<$1 Myr), but systematically overestimates the true pebble flux at late times ($>$1 Myr) by several orders of magnitude (Fig. \ref{fig:H2O_Mpeb}). The 1500/6000 K line ratio is shown to be sensitive to both the temperature and the dust distribution in the disk, and changes in this line ratio can often be induced by changes in the hot \ce{H2O} mass as well as the cold \ce{H2O} mass. 
\end{itemize}

This work demonstrates that modeling the transport of gas and dust to the inner disk in 2D brings interesting insights into the use of \ce{H2O} lines as a tracer of pebble drift. For future work, it will be important to fully consider the effects of transport on the entire \ce{H2O} reservoir, not just the cold reservoir, as this may have considerable effects on the observable tracers. Additionally, outward radial diffusion has been shown to be important for properly modeling the cold \ce{H2O} sublimating off the dust grains. Finally, it will be important to also consider the effects of gaps and dust traps, which may significantly alter the pebble flux and therefore can give further insights into the use of \ce{H2O} emission as a tracer of drift.


\begin{acknowledgements}
    Astrochemistry in Leiden is supported by funding from the European Research Council (ERC) under the European Union’s Horizon 2020 research and innovation programme (grant agreement No. 101019751 MOLDISK). E.v.D. also acknowledges support from the Danish National Research Foundation through the Center of Excellence ``InterCat'' (DNRF150).
\end{acknowledgements}

\bibliographystyle{aa}
\bibliography{references}

\begin{appendix}

\section{\ce{CO2}}\label{app:CO2}

\begin{figure*}
    \centering
    \includegraphics[width=0.85\linewidth]{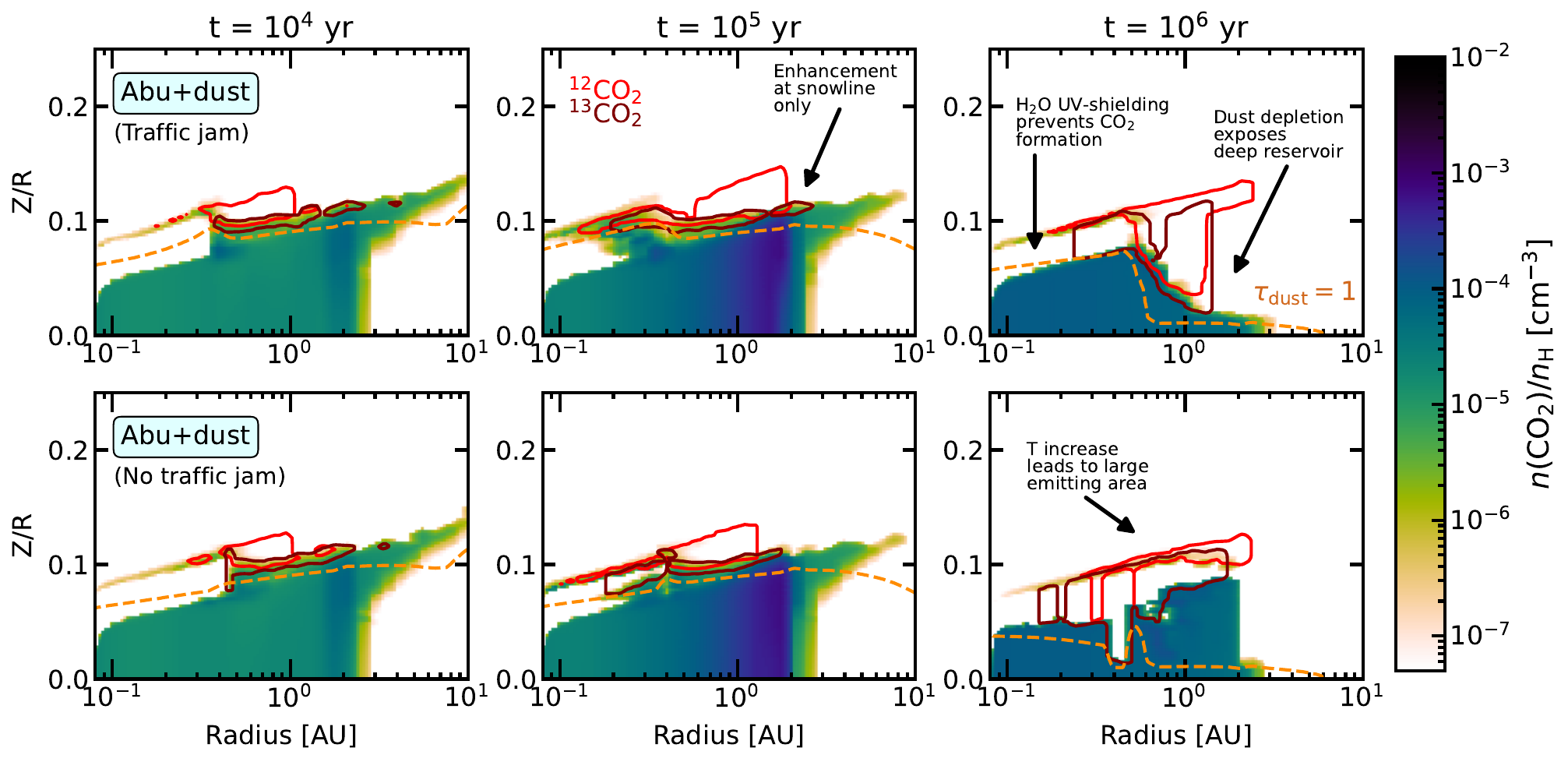}
    \caption{\ce{CO2} abundance as a function of radius and height at three time steps for the two grids with varied abundances and dust. In all panels, the dust $\tau=1$ surface at 15 $\mu$m is indicated with an orange dashed line. The red and brown contours in all panels represent the 80\% emitting regions of the \ce{^12CO2} and \ce{^13CO2} Q-branches, respectively.}
    \label{fig:CO2_abu}
\end{figure*}

\subsection{\ce{CO2} abundance and flux}\label{sub:co2_abu}

Fig. \ref{fig:CO2_abu} presents the gas-phase \ce{CO2} abundance throughout the disk in two of our five grids (the two grids with varied dust and abundances), shown at the same three time steps as Fig. \ref{fig:H2O_abu}. The full version with all five grids is presented in Fig. \ref{fig:CO2_abu_full}. As \ce{CO2} forms efficiently in the gas-phase at colder temperatures than \ce{H2O} ($\lesssim$300 K; \citealt{charnley1997, vandishoeck2013, walsh2015}), its gas-phase reservoir is located further out in the disk. However, in the warmer, inner regions of the disk, the presence of gas-phase \ce{H2O} can have a large impact on the \ce{CO2} abundance structure. When a sufficient column of \ce{H2O} is present ($\gtrsim10^{18}$ cm$^{-2}$; \citealt{bethell2009}), it can begin to self-shield against photodissociation. This quenches the formation of \ce{CO2}, as it relies on the presence of OH, a photodissociation product of \ce{H2O}. Simultaneously, the photodissociation rate of \ce{CO2} is not decreased by a similar amount, as no sufficient column is present to benefit from self-shielding or mutual shielding effects. Therefore, in regions where \ce{H2O} is abundant (around $Z/R\sim0.08$, within 0.5 au), one can observe a `hole' in the \ce{CO2} gas-phase abundance map \citep[see also][]{bosman2022b}. Above this hole, a thin layer of \ce{CO2} is present where a large-enough \ce{H2O} column has not yet been built up. This layer can actually contribute quite significantly to the total observed flux, especially when the temperature is high.

In terms of abundance variations with time, similar trends are observed as for the \ce{H2O} abundance (Fig. \ref{fig:H2O_abu}). A small build-up of gas-phase \ce{CO2} can be seen right at the midplane snowline (located at $\sim$2 au) at 10 kyr. At 0.1 Myr, this enhancement has increased and distributed itself further across the inner disk. Still, since the \ce{CO2} snowline is located further out than the \ce{H2O} snowline, this inwards advection occurs on a longer timescale for \ce{CO2}, and therefore has not yet reached the innermost regions of the disk. This only happens after $\sim$0.4 Myr (see Fig. \ref{fig:1Dcode_abu}). However, this enhancement in \ce{CO2} will also linger for a longer time. Therefore, by 1 Myr, \ce{CO2} is still enhanced, whereas \ce{H2O} shows a clear depletion. 

The impact of the dust properties is most clear at the latest times, when the dust has significantly depleted from the disk. The dust $\tau=1$ surface is located much deeper into the disk, especially in the models where no traffic jam is present (see orange dashed line). This exposes more of the colder, deeper-lying \ce{CO2} reservoir beyond 1 au than is otherwise typically visible.

\begin{figure*}
    \centering
    \includegraphics[width=\linewidth]{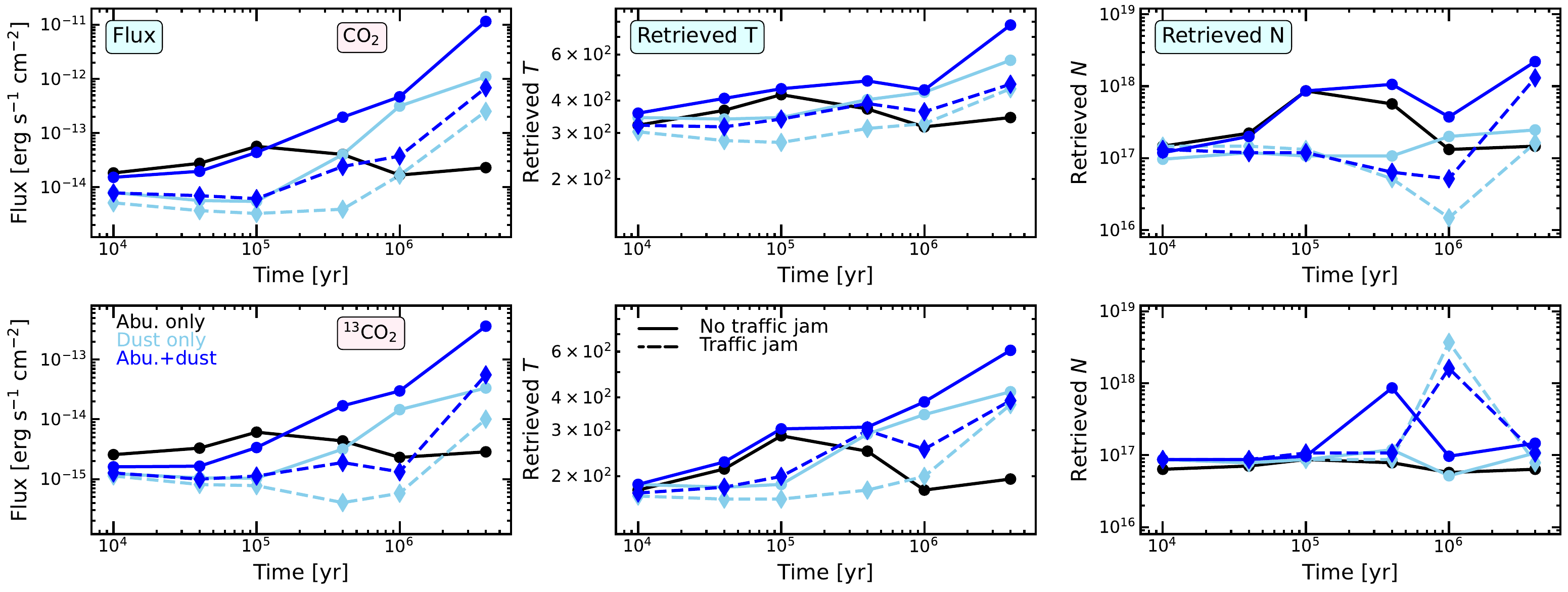}
    \caption{Left: integrated flux of the \ce{^12CO2} (top) and \ce{^13CO2} (bottom) Q-branches as a function of time for all grids. Middle: retrieved gas temperature from single-component LTE slab model fits. Right: retrieved column density from single-component LTE slab model fits. }
    \label{fig:CO2_cut}
\end{figure*}

The left column of Fig. \ref{fig:CO2_cut} presents the integrated fluxes of the \ce{^12CO2} and \ce{^13CO2} Q-branches as a function of time for all grids. Globally, the flux of both \ce{CO2} isotopologues closely follows the temperature evolution of the disk. When only the abundance is varied, a small temperature increase is present in the inner regions due to the influx of oxygen (see Fig. \ref{fig:tgas}). This temperature increase is clearly traced by an increase in \ce{CO2} flux. Though the \ce{CO2} also emits from further out in the disk ($\gtrsim$0.5 au) than where this temperature increase is present, it is the flux contribution from the thin, warm layer of \ce{CO2} within $\sim$0.2 au that is responsible for this flux increase. 

We note that this sensitivity of the \ce{CO2} emission to the disk temperature is very well traced by the shape of its $Q$-branch. This is demonstrated in Fig. \ref{fig:CO2_shape} for \ce{^12CO2}, but this is seen for all isotopologues. The leftmost panel of Fig. \ref{fig:CO2_shape} demonstrates how the $Q$-branch broadens around 0.1 Myr, and the peak shifts towards bluer wavelengths. This is a clear indication of an increase in temperature \citep[see, e.g.][]{cami2000, bosman2017}, as higher vibrational levels are becoming more excited. 

When only the dust properties are varied, the flux of all \ce{CO2} isotopologues remains relatively constant, until it strongly increases at late times when the dust depletes from the disk, and it starts to significantly heat up. This is also clearly visible in the shape of the $Q$-branch (Fig. \ref{fig:CO2_shape}). When a traffic jam is present, the increase in flux starts at later times, as the inner regions of the disk can remain cooler for longer in that scenario. However, whereas the warm \ce{H2O} flux showed a clear dip in flux around 0.1 Myr in this scenario due to the obscuration by dust (see left column, middle panel in Fig. \ref{fig:H2O_retrievedN}), this is not seen to the same extent in the \ce{CO2} flux. This is likely caused by the \ce{CO2} flux containing a contribution from regions which are either not affected by the dust pile-up (e.g., further out in the disk) and/or are not limited by the dust optical depth (see Sect. \ref{sub:co2_slabs}). When both the abundances and the dust are varied, a combined effect is seen in both flux and Q-branch shape, though the global trend is more similar to the simulations with varied dust only than to those with varied abundances only.

\subsection{Retrievals}\label{sub:co2_slabs}

The middle and right columns of Fig. \ref{fig:CO2_cut} present the results of LTE slab model fits to the emission of \ce{^12CO2} and \ce{^13CO2}. These fits are performed to the full emission features of these species, thus including their $P$-, $Q$-, and $R$-branches, as well as their hot bands. This once again demonstrates how the emission from all \ce{CO2} isotopologues is very sensitive to the temperature of the disk, as the retrieved temperatures very closely follow the previously identified trends that were also visible in the absolute integrated fluxes. The retrieved temperatures also demonstrate how \ce{^13CO2} traces slightly deeper and further-out regions in the disk (see also Fig. \ref{fig:CO2_abu}), as they trace colder temperatures. 

Unlike \ce{H2O}, however, the retrieved \ce{CO2} column density shows a less clear picture. In the case of \ce{^13CO2}, the emission is found to be (marginally) optically thin in all grids at almost all time steps. Therefore, the column density and emitting area are degenerate, and therefore the value of $\sim10^{17}$ cm$^{-2}$ that is found at most time steps represents an upper limit. At 1 Myr, in both models with a traffic jam present, the minor isotopologues show a spike in retrieved column density. This is most likely caused by the sharp drop in dust opacity across the \ce{H2O} snowline, that temporarily exposes a deeper-lying reservoir of \ce{CO2}, which preferentially enhances the column of the minor isotopologues. 

Interestingly, this is not present in the models without the traffic jam, despite the dust properties outside the \ce{H2O} snowline being the same in both models. However, the build-up of dust inside the \ce{H2O} snowline by the traffic jam also provides shielding of the radiation in the radial direction. Therefore, in the models without the traffic jam, there is a much stronger temperature increase in the inner disk, which leads to a much more extended \ce{H2O} gas reservoir (extending out to $\sim$2 au instead of 0.5, see Fig. \ref{fig:H2O_abu}). This \ce{H2O} provides shielding, allowing a larger midplane abundance of \ce{CO2} to build up around $\sim$2 au, and allowing the thin surface layer of \ce{CO2} to extend to the same radius. Therefore, most of the \ce{CO2} emission in the model without a traffic jam still comes from this thin layer, rather than the deeper-laying reservoir. In the model with the traffic jam, however, this thin surface layer is now much less radially extended, and a component from the deeper reservoir can be seen.

The emission from \ce{^12CO2} is slightly more sensitive to the total column than the isotopologues. When only the abundances are varied, an enhancement in retrieved column density is seen that corresponds well to the enhancement of gas-phase \ce{CO2} in the inner disk. When only the dust properties are varied, however, the retrieved \ce{CO2} column density generally remains roughly constant, and does not show the characteristic dip that is seen for \ce{H2O} when a traffic jam is present. Therefore, the \ce{CO2} column density seems rather insensitive to the properties of the dust. 

This can likely be explained by the \ce{CO2} abundance structure, and the origin of the emission: a significant fraction of the \ce{CO2} emission actually originates from the thin surface layer above the \ce{H2O} reservoir, rather than the more abundant, deeper-lying reservoir further out. The column of material present in this thin layer is set by the self-shielding of \ce{H2O}: once a large-enough \ce{H2O} column has been built up ($\sim10^{18}-10^{19}$ cm$^{-2}$; \citealt{bethell2009}), no more OH is available to form \ce{CO2}. Therefore, as long as the abundance of the constituent atoms (C and O) does not change (which holds true in the dust-only models), the column in this layer is going to be rather invariant to the dust properties of the disk, as the required column for \ce{H2O} self-shielding will generally be reached higher up in the disk than the $\tau=1$ surface. 

When both the dust and abundances are varied, the effects are mostly seen to combine again. Without a traffic jam, the increase in abundance is now traced again by the retrieved column. With a traffic jam, however, this is not the case, likely due to the $\tau=1$ layer being located much higher in the disk in the thin layer from which \ce{CO2} emits, therefore potentially hiding the enhancement.

\subsection{\ce{CO2}/\ce{H2O} ratios}\label{subsec:disc_co2}

\citet{houge2025_H2O} and \citet{sellek2025_CO2} predict based on 1D transport models that much of the delivery of gas-phase \ce{H2O} to the inner disk may go unnoticed due to the co-delivery of dust. This work reproduces this result. Fig. \ref{fig:H2O_retrievedN} demonstrates that the retrieved \ce{H2O} column density will indeed increase if only the gas-phase abundance in the inner region is increased (black lines), that it will decrease if only a dust pile-up in the inner region present (light blue dashed lines), and that the two effects cancel one another out almost entirely when both an enhanced gas-phase abundance \textit{and} a dust pile-up are present (dark blue dashed lines). Since the delivery of \ce{H2O} to the inner disk may thus go largely unnoticed, \citet{sellek2025_CO2} propose that the \ce{CO2}/\ce{H2O} column density ratio may provide a better tracer of pebble drift, as the emission from \ce{CO2} traces further out in the disk and is therefore not affected by the pile-up of dust to the same degree as \ce{H2O}. 


\begin{figure}
    \centering
    \includegraphics[width=\linewidth]{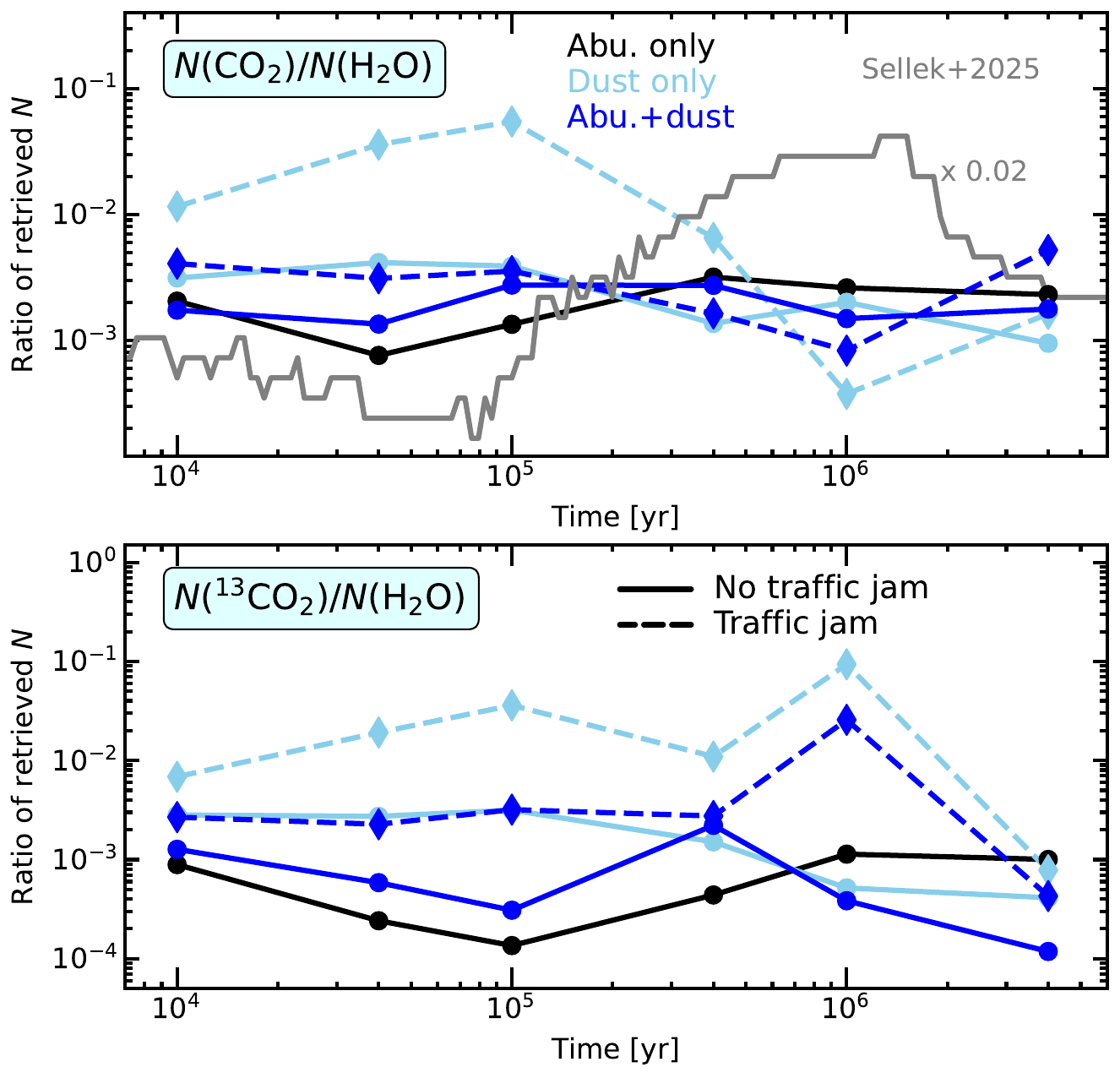}
    \caption{\ce{^12CO2}/\ce{H2O} and \ce{^13CO2}/\ce{H2O} retrieved column density ratios as a function of time for all grids.}
    \label{fig:ratio_retrievedN}
\end{figure}


Fig. \ref{fig:ratio_retrievedN} presents the ratio of the retrieved \ce{H2O} and \ce{CO2} column densities, where the warm \ce{H2O} column is used. In the grid with varied abundance only, the retrieved column density of both species traces its enhancement, and therefore the ratio nicely reflects how this happens in sequence: the \ce{H2O} abundance is enhanced at early times, leading to a decrease in the \ce{CO2/H2O} column density ratio, followed by an increase as the \ce{CO2} abundance is enhanced instead and the \ce{H2O} is already being drained onto the central star. This follows the behavior seen in Scenario 1 presented in \citet{sellek2025_CO2} (which has a traffic jam; see grey line in Fig. \ref{fig:ratio_retrievedN}), though the relative enhancement of \ce{CO2} with respect to \ce{H2O} is found to much more shallow in this work compared to \citet{sellek2025_CO2}. We also note that \citet{sellek2025_CO2} generally retrieve larger \ce{CO2} column densities, and therefore their \ce{CO2/H2O} column density ratio is scaled down by a factor 50 in Fig. \ref{fig:ratio_retrievedN} for easier comparison.

In the grids where only the dust is varied, however, a different behavior is seen. Now, the \ce{CO2} column density is found to be largely invariant to changes in the dust, so the behavior of the column density ratio is largely driven by \ce{H2O}. Therefore, when a traffic jam is present, an increase in the \ce{CO2/H2O} column density ratio is observed around 0.1 Myr due to the obscuration of \ce{H2O} by dust (dashed light blue line in Fig. \ref{fig:ratio_retrievedN}). When no traffic jam is present, the column density ratio instead remains roughly constant, perhaps showing a slight decrease with time. This may be due to the rising temperature in the disk at late times, which will allow the gas-phase \ce{H2O} reservoir to grow substantially, whereas the \ce{CO2} reservoir becomes smaller by comparison (see rightmost columns in Figs. \ref{fig:H2O_abu} and \ref{fig:CO2_abu}).

When the effects of both dust and gas transport are combined, the \ce{CO2/H2O} column density ratio remains relatively constant with time. With a traffic jam, now both the \ce{H2O} and \ce{CO2} columns are found to be roughly invariant with time, and without a traffic jam, both columns roughly trace a similar pattern of enhancement, and therefore mostly cancel out. Thus, unlike predicted by \citet{sellek2025_CO2}, the \ce{CO2/H2O} column density ratio does not seem more sensitive to the delivery of material than the \ce{H2O} column density by itself. 

Most of this can likely be attributed to the more complicated \ce{CO2} abundance structure that is used in this work compared to \citet{sellek2025_CO2}. Since a significant amount of \ce{CO2} emission traces the thin surface layer inside the \ce{H2O} snowline, its column is rather invariant and set by the \ce{H2O} shielding rather than the dust continuum. \citet{sellek2025_CO2}, on the other hand, assume a vertically constant abundance. Therefore, changes in the dust continuum do make a difference in the total \ce{CO2} column in their work, as the \ce{CO2} emission from inside the \ce{H2O} snowline becomes obscured, and the reservoir outside the snowline can start to dominate the \ce{CO2} emission instead. This effect is what allowed the \ce{CO2} emission to be sensitive to the delivery in their work, but since it is not present in ours, we do not find the same trend. 

Therefore, the \ce{CO2}/\ce{H2O} column density ratio may not provide a better tracer of drift. The interpretation of the \ce{CO2} emission is not straightforward, as the emission is quite sensitive to the disk temperature, and its abundance structure is complex. One may ask, however, whether it is realistic that the \ce{CO2} emission traces such a thin surface layer, as efficient vertical mixing may very rapidly spread it out, which is not accounted for in this work. The emission from \ce{^13CO2}, along with the other isotopologues, does emit from further out in the disk and therefore should be less affected by this invariance of its column to the dust continuum. 

However, the column density of any isotopologues will likely be poorly constrained due to the emission being optically thin. This is seen in the behavior of the \ce{^13CO2}/\ce{H2O} column density ratio (bottom panel of Fig. \ref{fig:ratio_retrievedN}), which generally reflects the inverse of the \ce{H2O} column density as the \ce{^13CO2} column density represents an upper limit for most models. Instead, one will need to consider the total mass or total number of molecules ($\mathcal{N} = N\times\pi R^2$), as this quantity will be properly constrained in this case. To fully understand what tracers may best reflect the delivery of material to the inner disk, further investigation beyond the scope of this paper will be needed.

    
    
    

\section{Supplementary figures}\label{app:figures}

\begin{figure*}
    \centering
    \includegraphics[width=\linewidth]{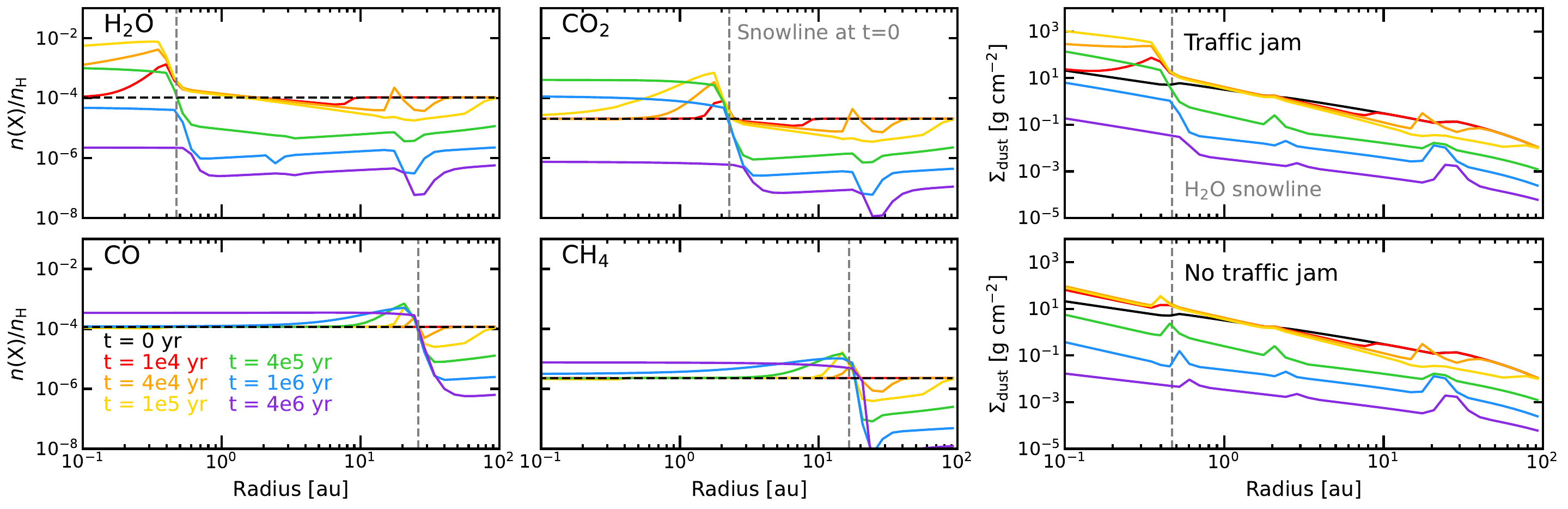}
    \caption{Left: Abundances of \ce{H2O, CO2, CH4}, and CO as a function of radius given by DiscEvolution at several time steps. The snowline of each species at $t=0$ is indicated by a gray dashed line. Right: Dust surface density as a function of radius at several time steps in the case with (top) and without (bottom) a traffic jam. The \ce{H2O} snowline at $t=0$ (interior to which the traffic jam occurs) is indicated with a gray dashed line. }
    \label{fig:1Dcode_abu}
\end{figure*}

\begin{figure*}
    \centering
    \includegraphics[width=0.9\linewidth]{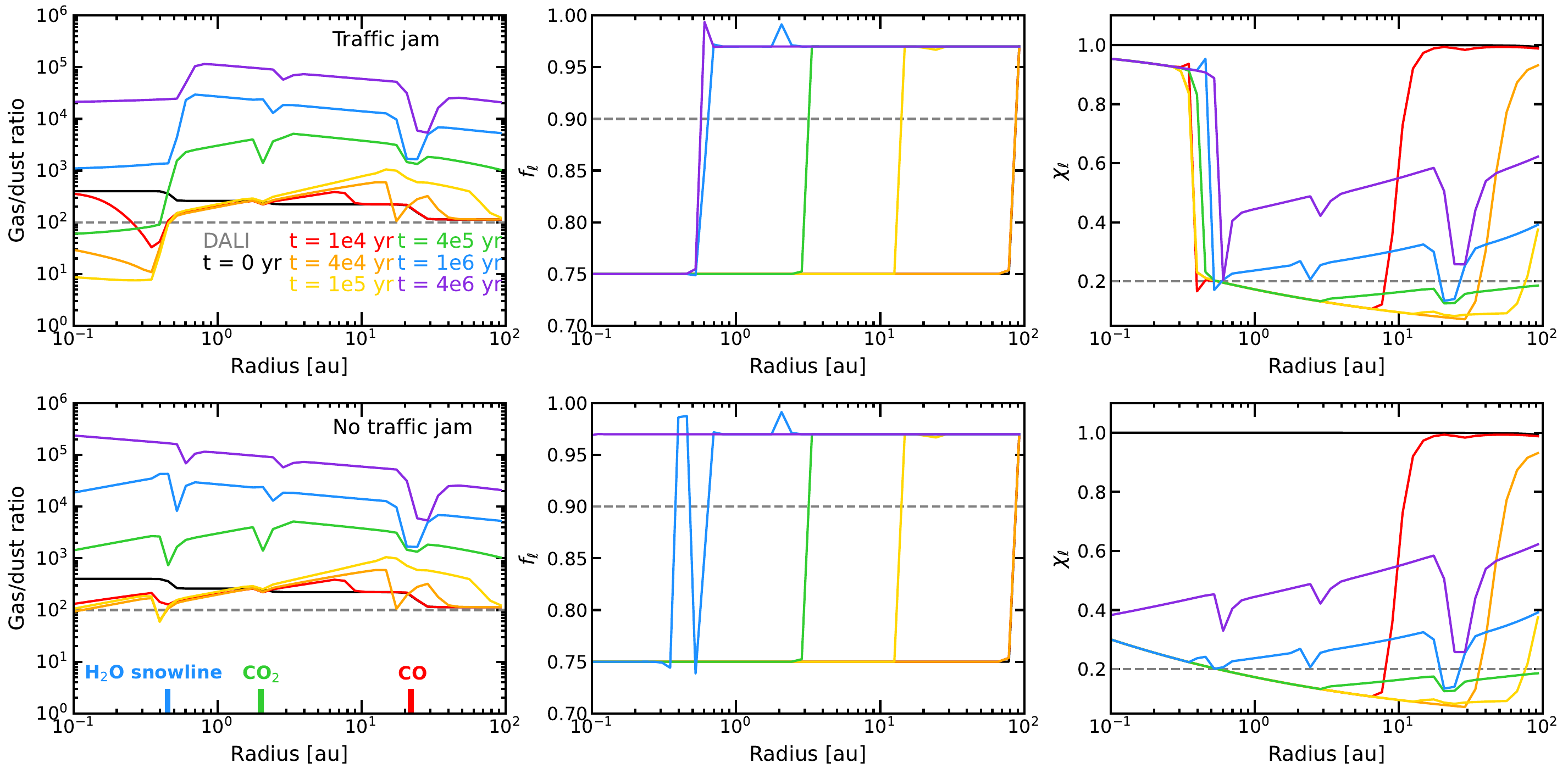}
    \caption{{Dust properties (vertically integrated gas/dust ratio $\Delta_{\rm gas/dust}$, mass fraction in large grains $f_{\rm large}$, and settling factor $\chi$) as a function of radius given by DiscEvolution at several time steps. The top row shows the scenario with a traffic jam ($u_{\rm f, dust}=1$ m s$^{-1}$, $u_{\rm f, ice}=10$ m s$^{-1}$), and the bottom row shows the scenario without a traffic jam ($u_{\rm f, dust}=u_{\rm f, ice}=10$ m s$^{-1}$). The gray dashed lines show the fiducial values used in the abundance-only grid. The dips are caused by the cold finger effect at the snowlines.} }
    \label{fig:1Dcode_dustprops}
\end{figure*}

\begin{figure*}
    \centering
    \includegraphics[width=\linewidth]{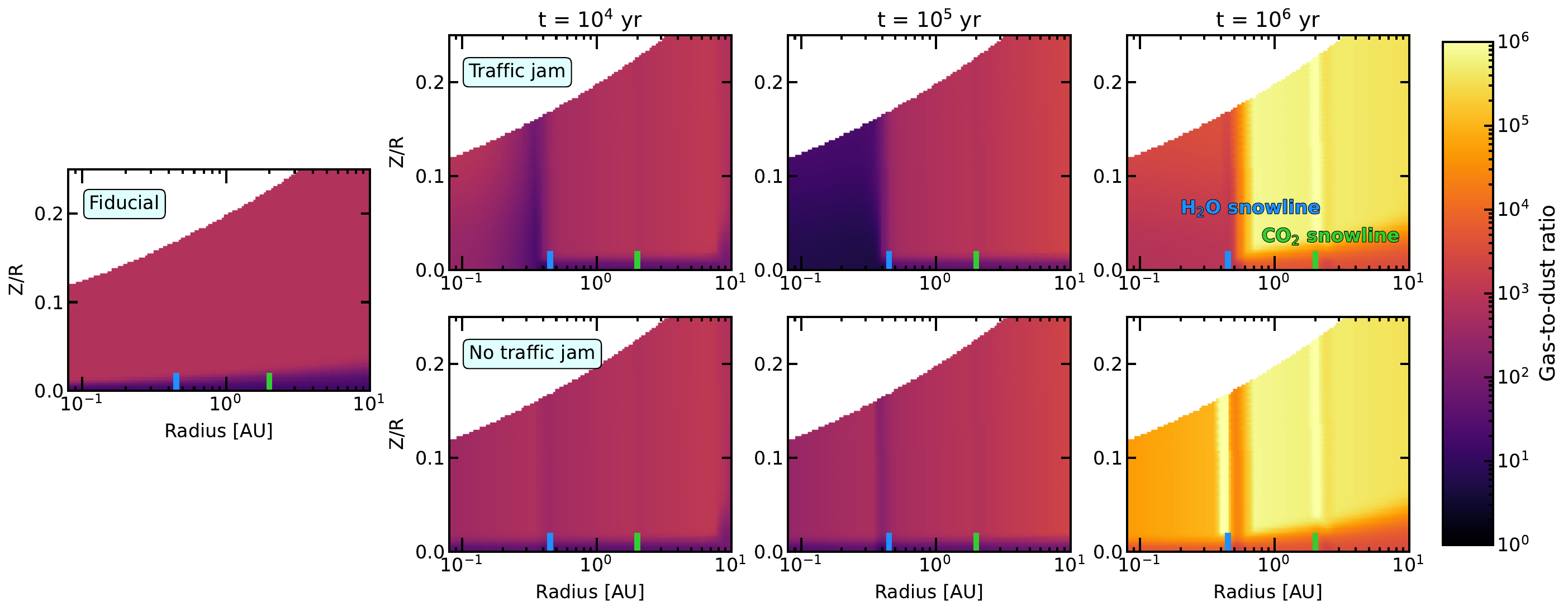}
    \caption{Gas-to-dust ratio given by DALI. The leftmost panel shows the fiducial distribution used by the grid with varied abundances only (see Table \ref{tab:grids}). The six remaining panels show the dust density created from the output of DiscEvolution at three different time steps, in both the scenarios with (top row) and without a traffic jam (bottom row).}
    \label{fig:gasdust}
\end{figure*}

\begin{figure*}
    \centering
    \includegraphics[width=\linewidth]{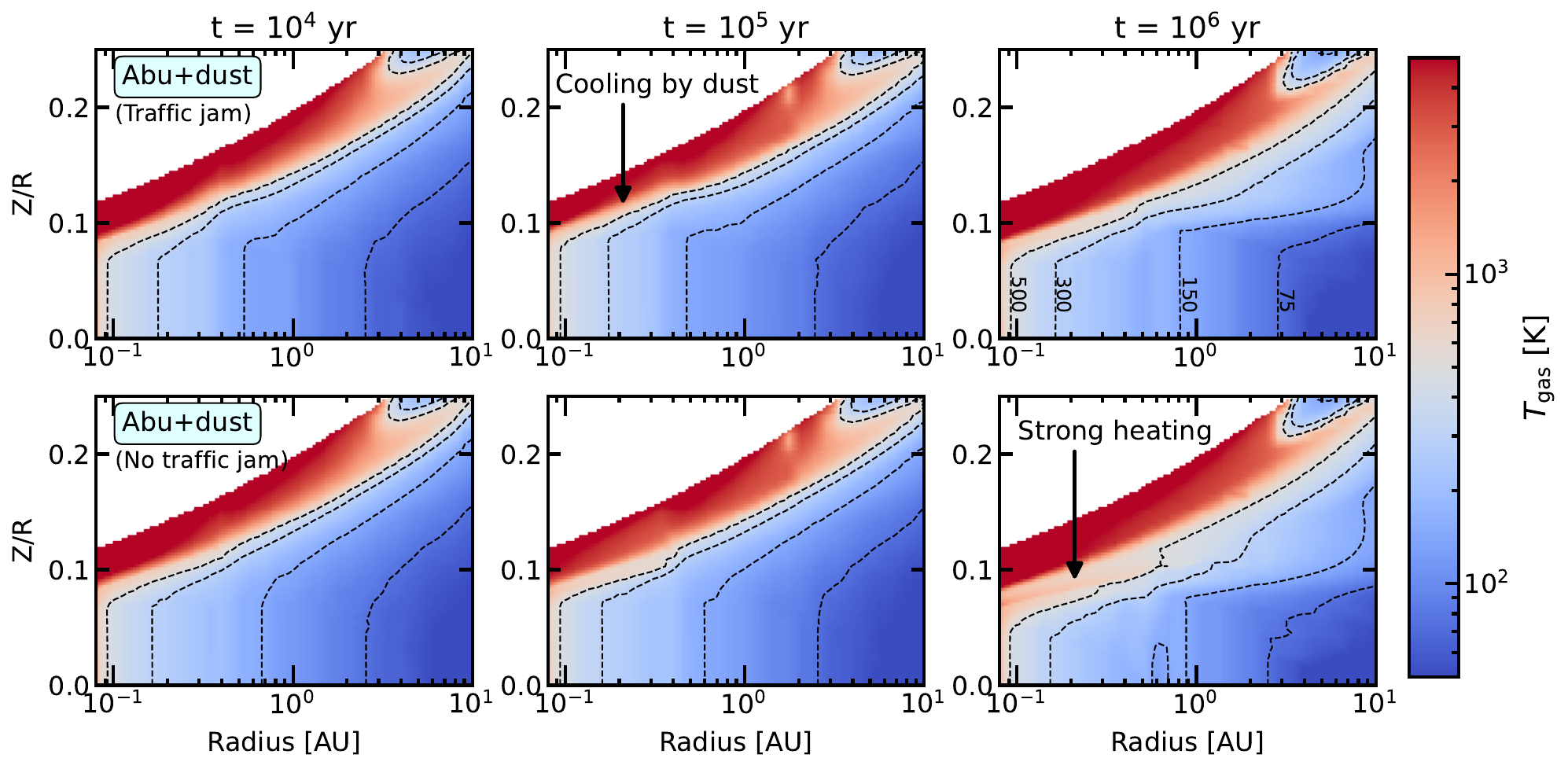}
    \caption{Gas temperature as a function of radius and height at three time steps for the two grids with varied abundances and dust. In all panels, temperature of 75, 150, 300, and 500 K are indicated with an black dashed lines.}
    \label{fig:tgas_map}
\end{figure*}

\begin{figure*}
    \centering
    \includegraphics[width=0.8\linewidth]{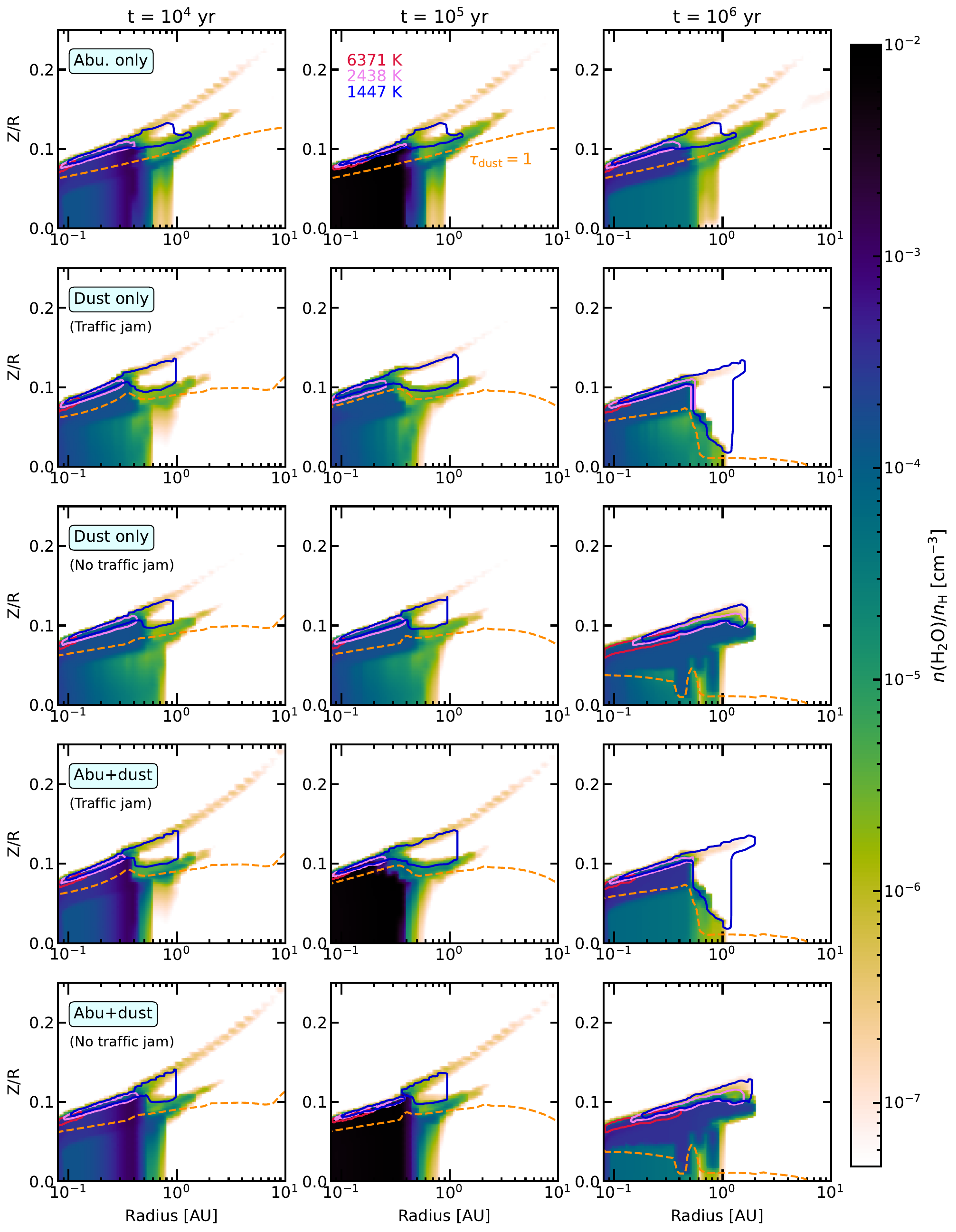}
    \caption{\ce{H2O} abundance as a function of radius and height at 3 time steps for all grids. In all panels, the dust $\tau=1$ surface at 15 $\mu$m is indicated with an orange dashed line. The red, pink, and blue contours in all panels represent the 80\% emitting regions of the \ce{H2O} 17$_{7,10}$ -- 16$_{4,13}$ ($E_{\rm up}$ = 6371 K), 11$_{3,9}$ -- 10$_{0,10}$ ($E_{\rm up}$ = 2438 K), and 8$_{3,6}$ -- 7$_{0,7}$ ($E_{\rm up}$ = 1447 K) lines, respectively, representing emission from hot, warm, and cold \ce{H2O}.}
    \label{fig:H2O_abu_full}
\end{figure*}

\begin{figure*}
    \centering
    \includegraphics[width=\linewidth]{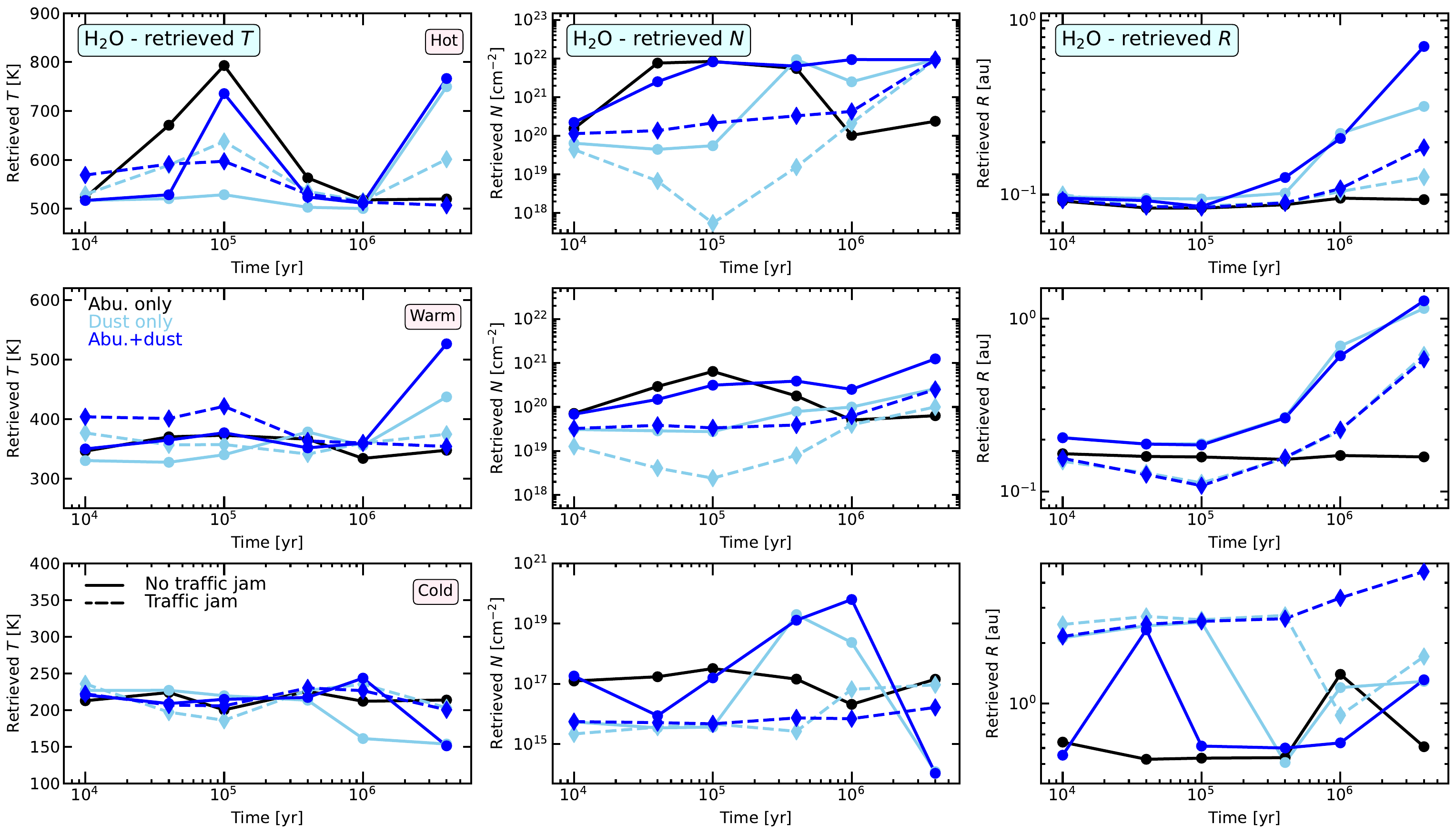}
    \caption{Retrieved \ce{H2O} gas temperature from single-temperature LTE slab model fits (left column) and a three-temperature-component MCMC fit (right column) as a function of time for all grids.}
    \label{fig:H2O_retrievedT}
\end{figure*}

\begin{figure*}
    \centering
    \includegraphics[width=0.8\linewidth]{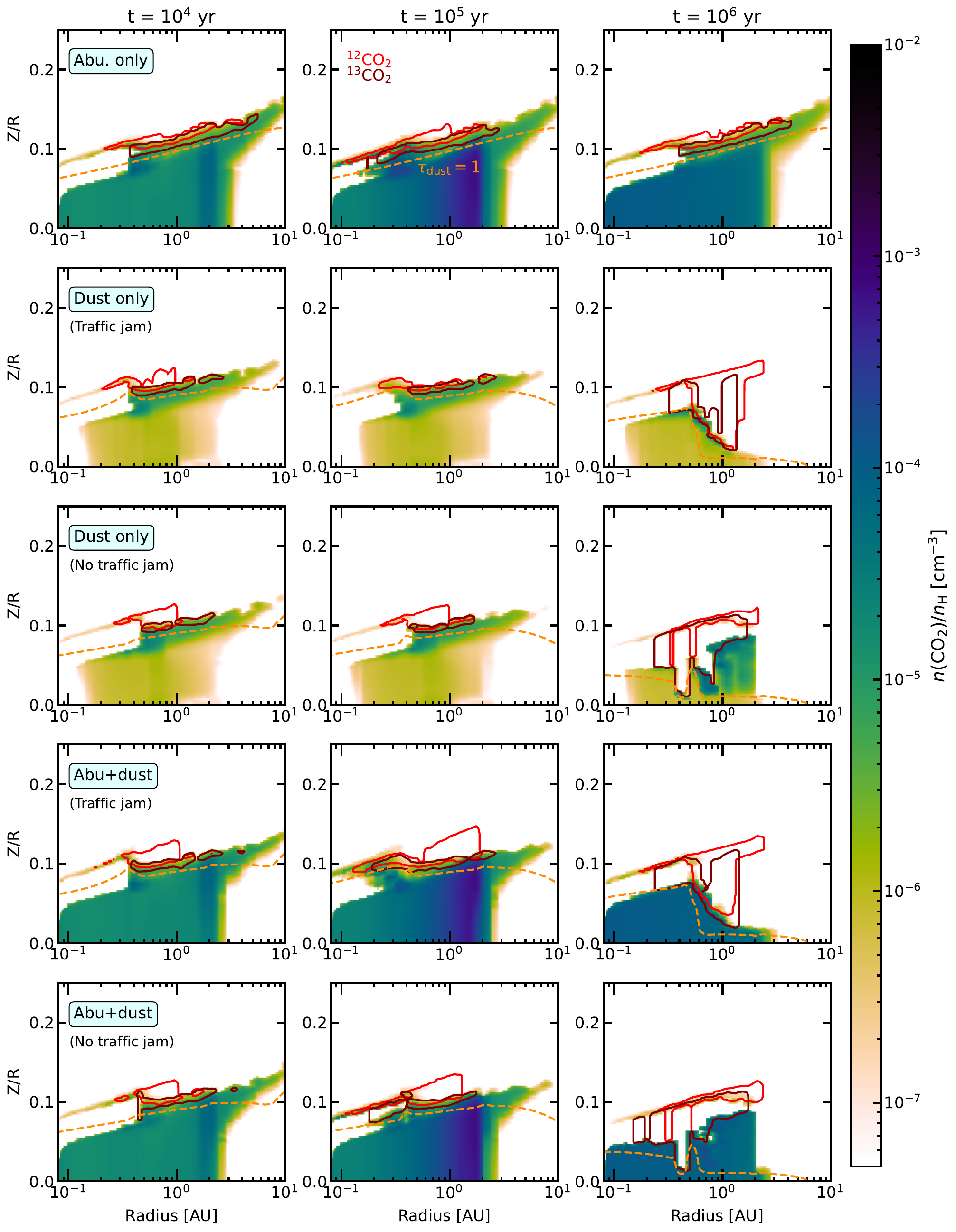}
    \caption{\ce{CO2} abundance as a function of radius and height at 3 time steps for all grids. In all panels, the dust $\tau=1$ surface at 15 $\mu$m is indicated with an orange dashed line. The red and brown contours in all panels represent the 80\% emitting regions of the \ce{^12CO2} and \ce{^13CO2} Q-branches, respectively. }
    \label{fig:CO2_abu_full}
\end{figure*}

\begin{figure*}
    \centering
    \includegraphics[width=\linewidth]{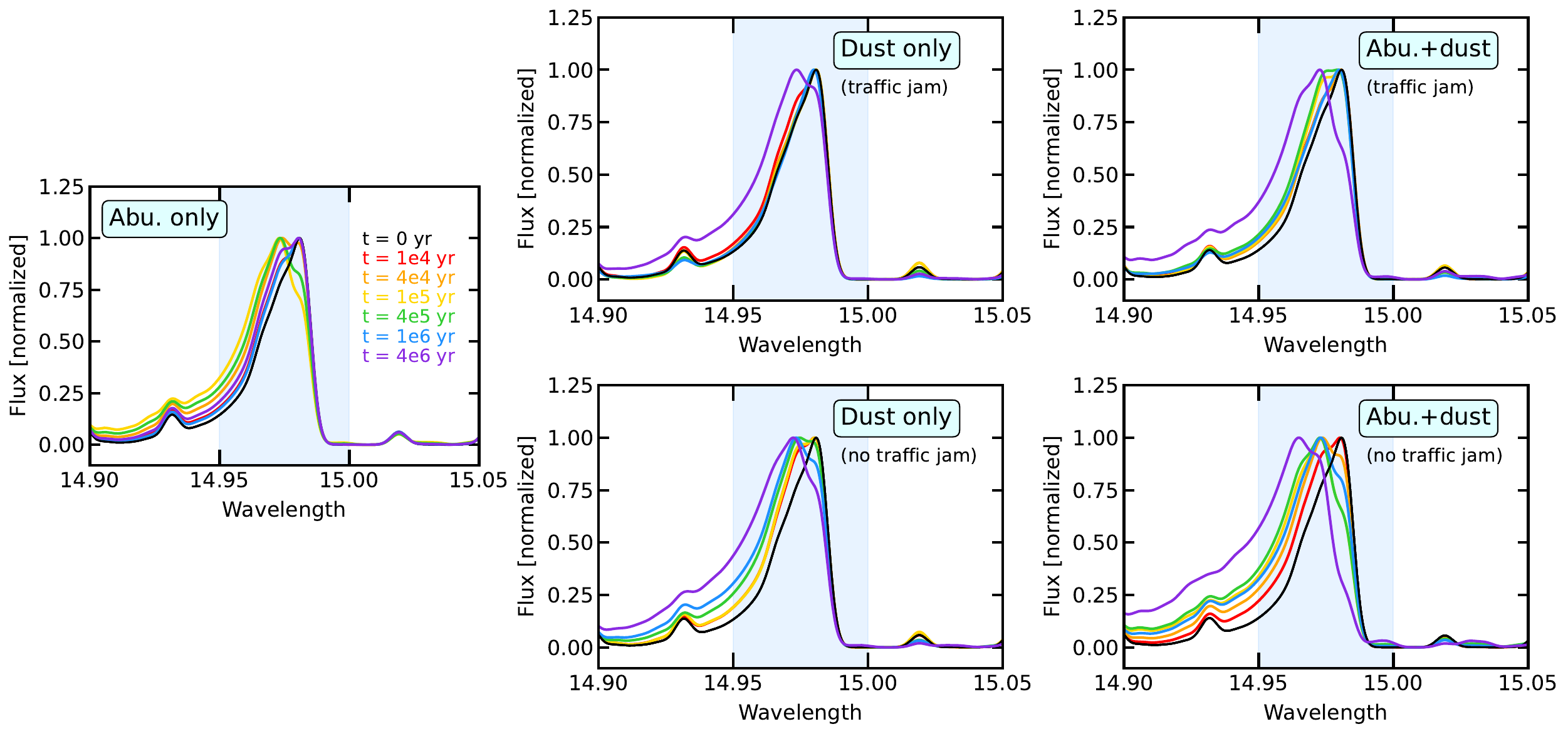}
    \caption{Synthetic \ce{^12CO2} spectra between 14.9 and 15.05 $\mu$m, normalized to the peak of the Q-branch (blue shaded region) for all grids at all time steps.  }
    \label{fig:CO2_shape}
\end{figure*}

Here, we present several supplementary figures. Fig. \ref{fig:1Dcode_abu} presents the gas-phase abundances of \ce{H2O}, \ce{CO2}, \ce{CO}, and \ce{CH4} predicted by DiscEvolution at several time steps. {Fig. \ref{fig:1Dcode_dustprops} presents the vertically integrated gas/dust ratio $\Delta_{\rm gas/dust}$, mass fraction in large grains $f_{\rm large}$, and settling factor $\chi$ as given by DiscEvolution at several time steps. }
Fig. \ref{fig:gasdust} presents the {2D map of the} gas-to-dust ratio {in the DALI models} throughout the disk at several time steps for both the fiducial case, as well as the cases with and without a traffic jam. Fig. \ref{fig:tgas_map} presents the full gas temperature maps for the two grids with varied abundances and dust. Fig. \ref{fig:H2O_abu_full} presents the full version of Fig. \ref{fig:H2O_abu}, containing the \ce{H2O} abundance for all five grids shown at three different time steps. Fig. \ref{fig:H2O_retrievedT} shows the full results of the MCMC retrievals on the \ce{H2O} emission, presenting the retrieved $T$, $N$, and $R$. Fig. \ref{fig:CO2_abu_full} presents the full version of Fig. \ref{fig:CO2_abu}, containing the \ce{CO2} abundance for all five grids. Fig. \ref{fig:CO2_shape} demonstrates how the $Q$-branch of the \ce{CO2} emission changes shape with the different time steps, therefore showing its sensitivity to the temperature of the disk.

\end{appendix}

\end{document}